\documentclass[a4paper,11pt]{article}
\pdfoutput=1
\usepackage{jheppub,MnSymbol}

\newcommand{\beq}{\begin{equation}}
\newcommand{\eeq}{\end{equation}}
\newcommand{\bea}{\begin{eqnarray}}
\newcommand{\eea}{\end{eqnarray}}

\preprint{MS-TP-17-09}

\title{Prompt photon production and photon-jet correlations at the LHC}

\author[a]{Michael Klasen,}
\author[b,c]{Christian Klein-B\"osing,}
\author[b]{Hendrik Poppenborg}

\affiliation[a]{Institut f\"ur Theoretische Physik, Westf\"alische Wilhelms-Universit\"at
 M\"unster, Wilhelm-Klemm-Stra\ss{}e 9, D-48149 M\"unster, Germany}
\affiliation[b]{Institut f\"ur Kernphysik, Westf\"alische Wilhelms-Universit\"at
 M\"unster, Wilhelm-Klemm-Stra\ss{}e 9, D-48149 M\"unster, Germany}
\affiliation[c]{ExtreMe Matter Institute EMMI, GSI Helmholtzzentrum f\"ur
 Schwerionenforschung, Planckstra\ss{}e 1, D-64291 Darmstadt, Germany}

\emailAdd{michael.klasen@uni-muenster.de}
\emailAdd{c.klein-boesing@uni-muenster.de}
\emailAdd{hendrik.poppenborg@uni-muenster.de}

\abstract{Next-to-leading order predictions matched to parton showers are compared
with recent ATLAS data on inclusive photon production and CMS data on associated
photon and jet production in pp and pPb collisions at different centre-of-mass
energies of the LHC. We find good agreement and, as expected, considerably reduced
scale uncertainties compared to previous theoretical calculations. Predictions are
made for the ratio of inclusive photons over decay photons $R_\gamma$, an important
quantity to evaluate the significance of additional photon sources, e.g.\ thermal
radiation from a Quark-Gluon-Plasma,
and for distributions in the parton momentum fraction in lead ions $x_{\rm Pb}^{\rm obs}$,
that could be determined by ALICE, ATLAS, CMS and LHCb in ongoing analyses of photon+jet
production in pPb collisions at $\sqrt{s_{NN}}=5.02$ TeV. These data should have an
important impact on the determination of nuclear effects such as shadowing at low $x$.}

\keywords{Perturbative QCD, NLO, parton showers, photons, hadron colliders}

\begin{document}
\maketitle
\flushbottom

%%%%%%%%%%%%%% Begin Section 1 %%%%%%%%%%%%%%%%%%%%%%%%%%%%%%%%%%%%%%%%%
\section{Introduction}
\label{sec:1}

Photons play many important roles in high-energy collisions. They allow, e.g.,
to determine the strong coupling constant ($\alpha_s$) \cite{Albino:2002ck} or
the partonic structure of protons, photons \cite{Klasen:2002xb} and nuclei
\cite{Stavreva:2010mw}, and they can serve as a thermometer of the hot
Quark-Gluon-Plasma (QGP) created in heavy-ion collisions \cite{Arnold:2001ms,%
Klasen:2013mga}.
If produced in association with a hadronic jet, they allow to calibrate the
transverse momentum of the latter at its creation and to deduce its subsequent
hadronic energy loss or azimuthal decorrelation in the medium \cite{Wang:1996yh}.

Theoretical predictions for final-state photon production in pp, pPb, and PbPb
collisions have since long been available at leading order (LO) of perturbative
QCD and have been implemented in parton shower (PS) Monte Carlo generators like
PYTHIA 8 \cite{Sjostrand:2007gs}. The latter also allow to model the (vacuum)
fragmentation
of final-state quarks and gluons into the observed hadrons and jets. Here,
detailed information on the final state comes at the price of large theoretical
uncertainties from variations of unphysical scales and other parameters.
Nevertheless, Monte Carlo generators represent an indispensable tool in the
experimental analyses such as those of the ALICE \cite{ALICE:2014rma,Adam:2015lda},
ATLAS \cite{Aaboud:2017cbm,Aad:2015lcb}, CMS \cite{CMS:2013oua}, and LHCb
\cite{Aaij:2016ofv} experiments at the Large Hadron Collider (LHC) at CERN.

In contrast, inclusive next-to-leading order (NLO) QCD calculations
\cite{Catani:2002ny} depend on
fewer tunable parameters and allow for a stabilisation of renormalisation and
factorisation scale dependences among perturbative matrix elements and
renormalisation-group improved initial-state parton density functions (PDFs)
and final-state fragmentation functions (FFs).
%++
As a consequence, they have been widely applied to photon measurements
at the Tevatron, most recently to single \cite{Aaltonen:2009ty} and pairs of
isolated photons \cite{Abazov:2013pua}, photon+jet \cite{D0:2013lra}, and
(with shortcomings at high $p_T$) photon+heavy quark measurements
\cite{Aaltonen:2013ama}.
%--
Unfortunately, PDFs and FFs
are non-perturbative and must be fitted to experimental data \cite{Bourhis:1997yu}.
Due to the lack of a new electron-positron collider, few improvements have
been made on photon FFs over the last decades \cite{Klasen:2014xfa}.
%++
NLO calculations for $Z \gamma$+jet and Z$\gamma \gamma$ \cite{Campbell:2012ft}
have recently become available in MCFM, followed by predictions at
next-to-next-to-leading order (NNLO) for diphoton \cite{Campbell:2016yrh} and
$Z\gamma$ \cite{Campbell:2017aul} production.
%--
The disadvantages of inclusive calculations lie in the limited multiplicity
of and information on the produced final state and the consequently restricted
number of kinematic observables.

Matching a NLO calculation with a PS Monte Carlo generator
combines the advantages of both methods. Subtraction methods \cite{Frixione:1995ms}
like the one implemented in POWHEG \cite{Frixione:2007vw} to avoid
double counting of the soft and collinear regions are now well established.
We have recently applied this method to prompt photon production and
associated photon-jet production in proton-proton (pp) collisions and
compared our numerical results with PHENIX data from RHIC at BNL
\cite{Jezo:2016ypn}. An important difference of NLO+PS calculations
with respect to inclusive NLO calculations lies in the treatment
of fragmentation contributions. It has, however, been demonstrated
that in electron-positron collisions the NLO+PS approach also leads
to a very good description of FF photons \cite{Hoeche:2009xc}.
%++
Multijet merging strategies have also recently been developed and
applied in particular to diphoton production at NLO
\cite{Gehrmann:2013bga} and at NLO+PS \cite{Siegert:2016bre}.
%--

It is the purpose of this paper to confront our NLO+PS calculations
employing POW\-HEG+PYTHIA for the first time with LHC data. After a review
of our theoretical method and a demonstration of the parameter
dependence of LO+PS calculations in Sec.\ \ref{sec:2}, we perform
these comparisons in Sec.\ \ref{sec:3} first for inclusive photon
production, measured in pp collisions by ATLAS \cite{Aaboud:2017cbm},
then for associated photon+jet production, analysed in pp and pPb
collisions by CMS \cite{CMS:2013oua}. Predictions for these experiments
as well as for the ALICE and LHCb experiments are made in Sec.\ \ref{sec:4} with a
focus on the determination of nuclear PDFs in pPb collisions and on
the reliable calculation of quantities relevant for the extraction
of QGP properties. Our conclusions are presented in Sec.\ \ref{sec:5}.

%%%%%%%%%%%%%% Begin Section 2 %%%%%%%%%%%%%%%%%%%%%%%%%%%%%%%%%%%%%%%%%
\section{Prompt photon production with POWHEG}
\label{sec:2}

In this section, we summarise our theoretical calculation of prompt photon
production at NLO and the implementation of this calculation in POWHEG.
Details can be found in our previous publication \cite{Jezo:2016ypn}.

\subsection{NLO calculation of prompt photon production and matching to PS}

Direct photon production proceeds at LO through the tree-level processes
$q\bar{q}\to\gamma g$ and $qg\to\gamma q$. In the POWHEG method \cite{Frixione:2007vw}, these
processes must also be calculated with colour and spin correlations. We
computed the traces of Dirac matrices with FormCalc 8.4 \cite{Fritzsche:2013fta}
and checked our results against the literature and Madgraph 5 \cite{Alwall:2011uj}.

The virtual corrections required the computation of the same processes at the one-loop
level, the subsequent reduction of tensor loop integrals with Form \cite{Vermaseren:2000nd},
and the computation of the resulting scalar integrals with LoopTools 2.13
\cite{Hahn:1998yk}. Renormalisation was performed in the $\overline{\rm MS}$
scheme, and the UV-finite results were checked against MG5$\_$aMC@NLO
\cite{Alwall:2014hca}.
For the real corrections, the tree-level processes with additional gluon
radiation or gluon splitting into quark-antiquark pairs had to be computed.
The new process $gg\to\gamma q\bar{q}$ plays a particularly important role
at the LHC due to the high gluon luminosity at these collision energies.
Final-state
quark-antiquark pairs lead to different jet fragmentation, in particular
in the QGP produced in heavy-ion collisions. Again, traces of Dirac matrices
were computed with FormCalc 8.4 and the results checked with Madgraph 5.
Infrared singularities were then removed with the dipole subtraction method
\cite{Catani:1996vz}. QCD finiteness was checked against AutoDipole 1.2.3
\cite{Hasegawa:2009tx}, and the integrated QED dipole was shown to reproduce
well the pointlike contribution to the photon fragmentation function.
The total NLO calculation was finally shown
to agree with JETPHOX \cite{Catani:2002ny}. The latter, however, also
includes fragmentation contributions at NLO and requires the convolution with
a photon fragmentation function such as BFG II \cite{Bourhis:1997yu}.
In our POWHEG approach, these contributions are taken into account at NLO
by partonic scatterings supplemented with a QED PS.
For our central numerical results, we identified the renormalisation and
factorisation scales
%++
$\mu_R$ and $\mu_F$
%--
with the transverse momentum of the photon or leading
underlying parton. To estimate the theoretical uncertainties, the scales
were then varied by relative factors of two, but not four.

NLO calculations usually increase and, more importantly, stabilise the LO
cross section with respect to scale variations. However, they include only
one additional parton and no hadronisation effects. Conversely, PS Monte
Carlo generators generally have LO normalisation and large scale dependence,
but include many additional partons and different hadronisation models.
Matching NLO with PS combines the advantages of both approaches. In POWHEG,
the overlap in the soft/collinear regions of real emission processes is
automatically subtracted with the FKS method \cite{Frixione:1995ms}, so that
only finite (i.e.\ UV-renormalised and IR-subtracted) loop and unsubtracted
real emission amplitudes must be provided. The hardest radiation
is generated first, only Monte Carlo events with positive weight are
produced, and the NLO calculation can be combined with various PS algorithms.
In our work we use PYTHIA 8.226 \cite{Sjostrand:2007gs} %.
%++
with its $p_T$-ordered parton shower and string hadronisation model
and allow unstable short-lived particles to decay.
We have checked that multiple interactions and soft underlying events, while
included in our calculations, have no significant effects.
To account for different definitions of radiation transverse momentum in POWHEG
BOX and PYTHIA, we use the class PowhegHooks and have PYTHIA evolve the shower
starting from the kinematical limit rather than the scale passed by the POWHEG BOX.
We then translate the $p_T$ of a generated radiation from the PYTHIA definition to
the POWHEG BOX definition and veto radiation harder than the hard POWHEG scale.
Following Ref.\ \cite{DErrico:2011cgc}, we employ different hard scales for the
QED and for the QCD shower. In the nomenclature of Ref.\ \cite{Barze:2014zba}, we
use the so-called NC-scheme and modify the POWHEG BOX to pass the scales of
both the underlying Born event and the first radiation to PYTHIA. For completeness,
we list non-standard POWHEG parameter settings and options in Tabs.\ \ref{tab:1}
and \ref{tab:2}.
\begin{table}
\caption{\label{tab:1}List of POWHEG BOX parameters set to different values than those in the manual.}
\vspace*{2mm}
\begin{tabular}{|c|c|c|c|c|c|c|c|c|c|}
\hline
ncall1 & \!itmx1\! & ncall2 & \!itmx2\! & \!foldcsi\! & \!foldy\! & nubound & \!iymax\! & doublefsr & iupperfsr \\
\hline
\!500000\! & 5     & \!500000\! & 5     & 2       & 5     & \!500000\!  & 5     & 1         & 1         \\
\hline
\end{tabular}
\end{table}
\begin{table}
\caption{\label{tab:2}List of POWHEGHooks options.}
\vspace*{2mm}
\centering
\begin{tabular}{|c|c|c|c|c|c|}
\hline
veto & vetoCount & pThard & pTemt & pTdef & QEDveto \\
\hline
1    & 50        & 0      & 0     & 1     & 0       \\
\hline
\end{tabular}
\end{table}
%
%--

\subsection{Sensitivity of the QED PS to the cutoff scale}

An important difference between the inclusive fixed-order and the more
differential PS calculations is the treatment of fragmentation contributions.
At fixed order, infrared singularities are consistently factorised into
non-perturbative fragmentation functions, whereas they are simply cut off in
the PS by allowing it to evolve in virtuality, angle, or transverse momentum
in each subsequent branching down to an arbitrary cut-off scale $Q_0^2$. The
maximal allowed $Q^2$ is set by the hard scattering process.

\begin{figure}[!h]
 \centering
 \includegraphics[width=0.49\linewidth]{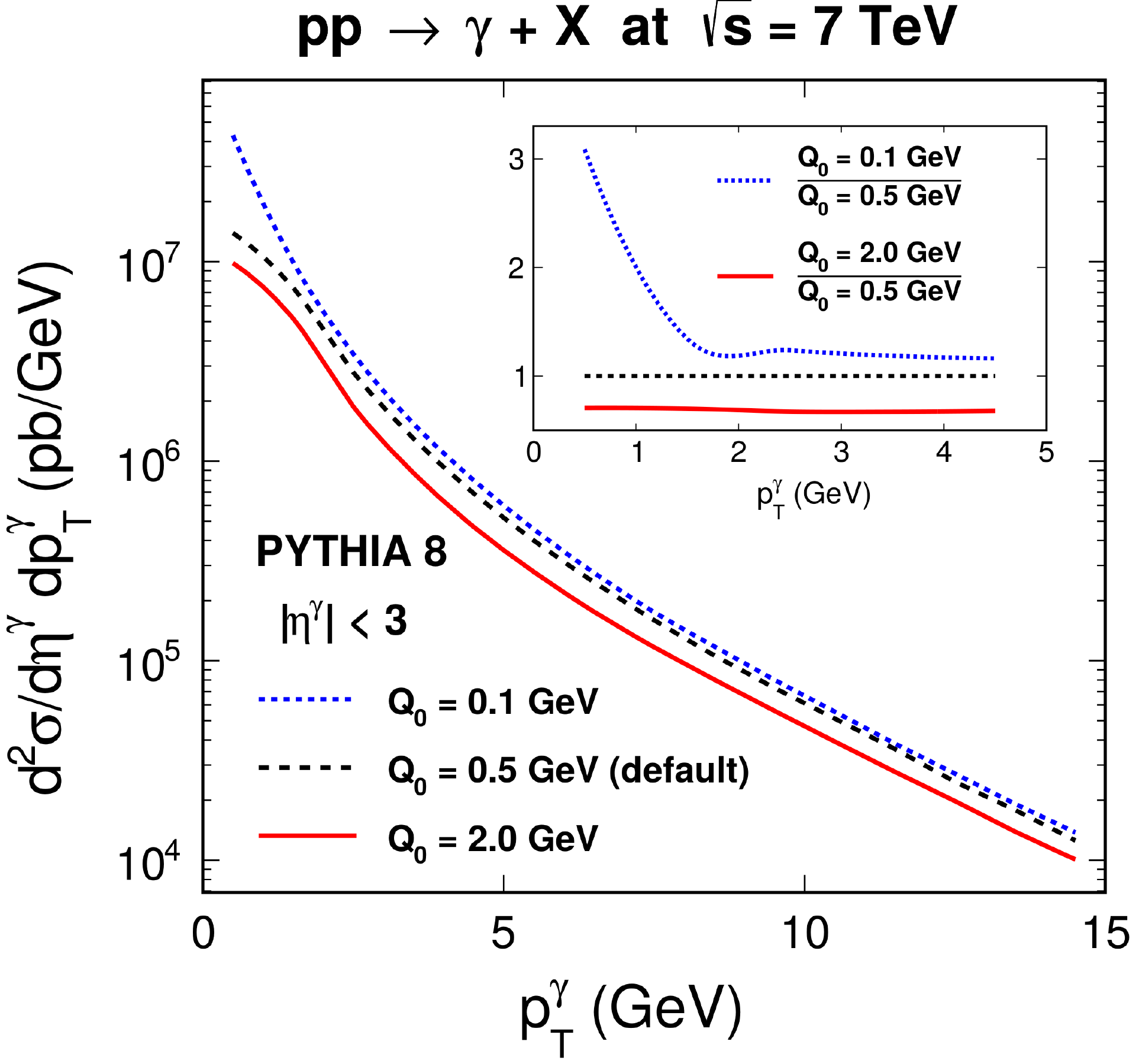}
 \caption{\label{fig:11}Transverse momentum dependence of inclusive photons
 produced in pp collisions at the LHC with a centre-of-mass energy of $\sqrt{s}=7$
 TeV for different PS cut-off scales $Q_0$ in PYTHIA 8.}
\end{figure}

The transverse momentum dependence of inclusive photons generated by the
QED PS in the Monash 2013 tune of PYTHIA 8 is shown in Fig.\ \ref{fig:11}.
The photons are assumed to be produced in pp collisions at the LHC with a
centre-of-mass energy of $\sqrt{s}=7$ TeV and detected in the CMS electromagnetic
calorimeter with rapidity $|\eta^\gamma|<3$. No isolation criterion on the photons
is applied. As in the Monash 2013 tune, the PDF fit NNPDF2.3QED LO with
$\alpha_s(M_Z)=0.130$ is used. The dependence of the transverse momentum
spectrum on the lower cut-off scale $Q_0$ can be deduced from the different
curves and from the ratios of the variations $Q_0=0.1$ GeV (dotted blue) and 2 GeV
(full red) to the default choice 0.5 GeV (dashed black) in the inset plot. Above
transverse photon momenta of $p_T^\gamma=2$ GeV, the results with $Q_0=0.1$ GeV and
$Q_0=2$ GeV differ from those at 0.5 GeV only by a constant factor of $+10$\% and
$-20$\%, respectively. Towards low $p_T^\gamma$, the results with $Q_0=0.1$ GeV
rise, however, much more steeply than those with $Q_0=0.5$ GeV.
These results demonstrate the strong dependence on and the need for tuning
of the arbitrary cut-off $Q_0$ in a LO PS Monte Carlo generator. In contrast,
the POWHEG approach guarantees for a correct cancellation of infrared
singularities among NLO matrix elements and PS.

%%%%%%%%%%%%%% Begin Section 3 %%%%%%%%%%%%%%%%%%%%%%%%%%%%%%%%%%%%%%%%%
\section{Comparisons with current LHC data}
\label{sec:3}

We now turn to comparisons of our NLO+PS predictions with POWHEG+PYTHIA
with published LHC data. For comparison, we will also show calculations
at LO+PS with PYTHIA 8 and at NLO with JETPHOX, since these were so far
frequently used as theoretical baselines in the experimental publications.

\subsection{Inclusive photon production in pp collisions with ATLAS}

The ATLAS collaboration has measured the production of inclusive photons
in pp collisions at the LHC with a centre-of-mass energy of $\sqrt{s}=13$
TeV and an integrated luminosity of 3.2 fb$^{-1}$ accumulated in 2015
\cite{Aaboud:2017cbm}.
To avoid the large contribution of photons from decays of energetic neutral
pions and $\eta$-mesons inside jets, the photons were isolated based on the
amount of transverse energy $E_T^{\rm iso}$ inside a cone of size $\Delta R=
\sqrt{(\Delta\eta)^2+(\Delta\phi)^2}=0.4$ in the plane of rapidity ($\eta$)
and azimuthal angle ($\phi$), excluding an area of size $\Delta\eta \times
\Delta\phi = 0.125\times 0.175$, around the photon candidate. Specifically,
$E_T^{\rm iso}$ was required to be lower than $4.8+4.2\cdot10^{-3}\cdot E_T^\gamma$
[GeV]. 

\begin{figure}[!h]
 \includegraphics[width=\linewidth]{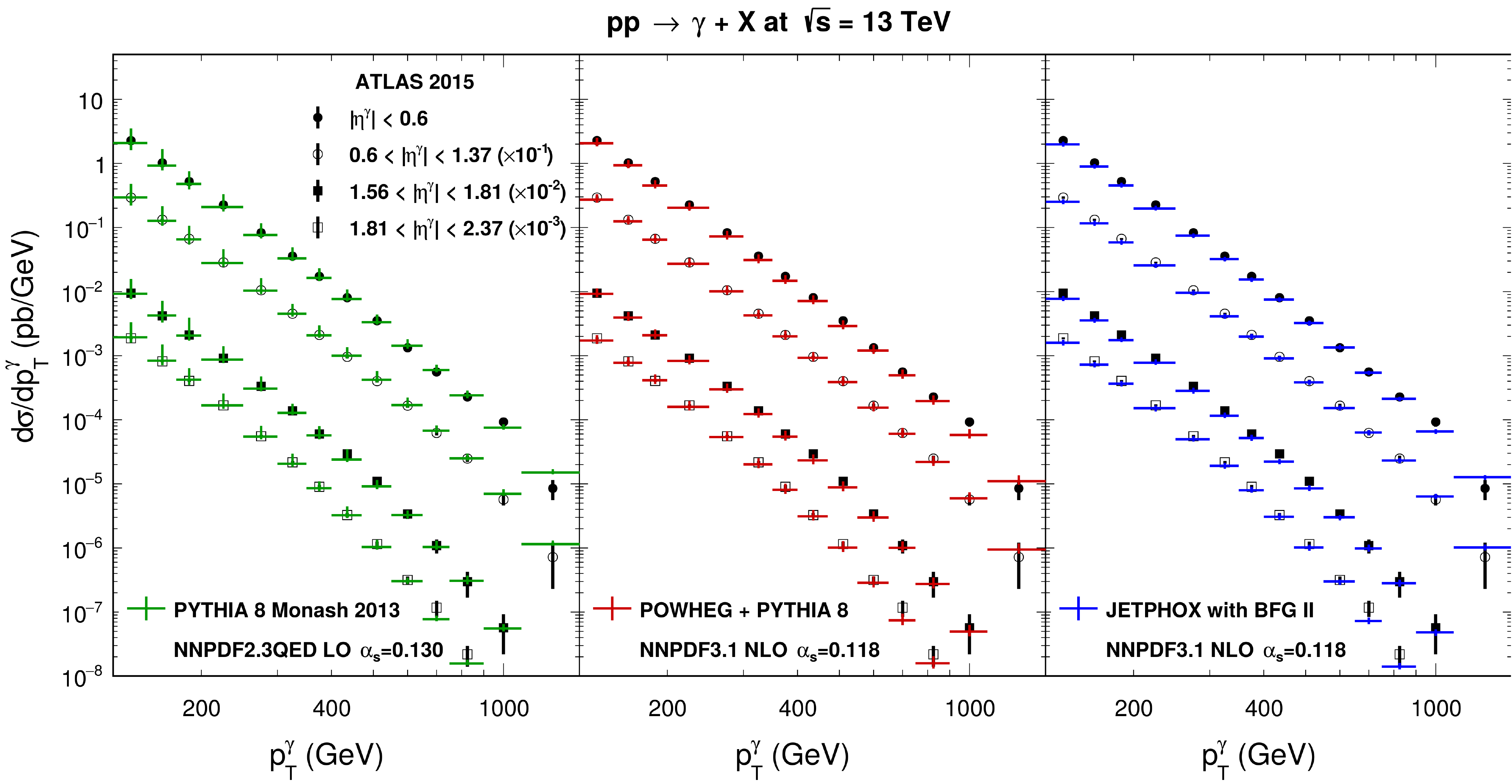}
 \caption{\label{fig:1}Transverse momentum distributions of isolated photons in
 pp collisions at the LHC with a centre-of-mass energy of $\sqrt{s}=13$ TeV in
 four different rapidity bins. ATLAS 2015 data are compared to predictions in
 LO with PYTHIA 8 (left), NLO with JETPHOX (right), and NLO+PS with POWHEG
 (centre).}
\end{figure}

In Fig.\ \ref{fig:1}, we compare the transverse momentum distributions
measured by ATLAS in four rapidity ranges to predictions in LO from the
PYTHIA 8 PS Monte Carlo generator (green, left) \cite{Sjostrand:2007gs}, in NLO
from the fixed-order program JETPHOX \cite{Catani:2002ny} using the BFG
II photon fragmentation function (blue, right) \cite{Bourhis:1997yu}, and to our new
NLO+PS predictions with POWHEG matched to PYTHIA 8 (red, centre) \cite{Jezo:2016ypn}.
For the NLO and NLO+PS predictions, we employ the recent PDF set NNPDF3.1
NLO based on high-precision collider data \cite{Ball:2017nwa}. 
For the LO predictions, we use instead the preceding set NNPDF2.3QED LO
\cite{Ball:2013hta}, which (while based on a similar theoretical approach) was
in particular employed in the Monash 2013 tune of PYTHIA 8 \cite{Skands:2014pea}.
For inclusive photon production, on a logarithmic scale, and thanks to the tuning of
the Monte Carlo generator, all three predictions seem to agree very well with
the data. Smaller discrepancies are, however, already visible in LO and NLO at high
transverse momenta and/or very forward rapidities, while the NLO+PS predictions
always agree with the data within the experimental error bars.

\begin{figure}[!h]
 \includegraphics[width=\linewidth]{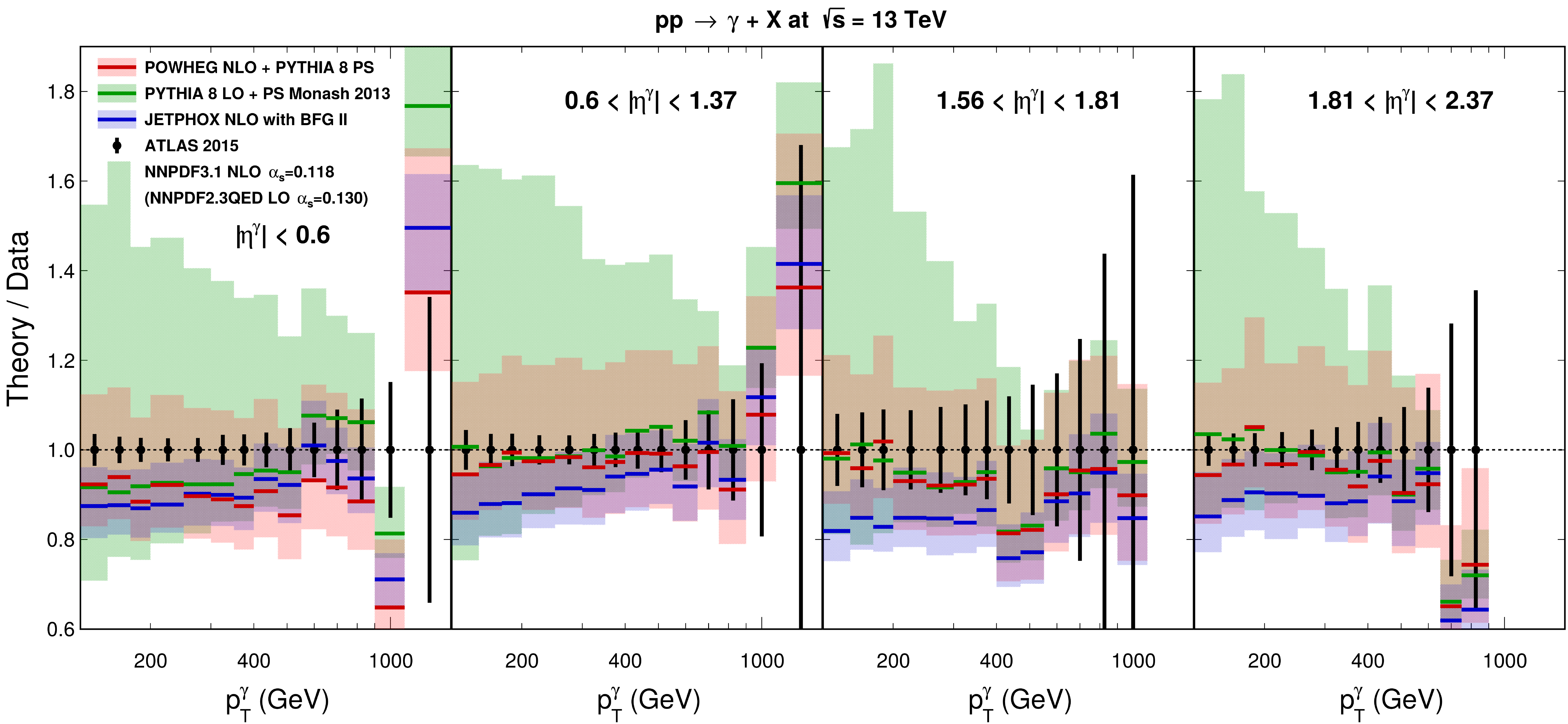}
 \caption{\label{fig:2}Same as Fig.\ \ref{fig:1}, but with theory normalised
 to data and including scale uncertainties.}
\end{figure}

A more detailed analysis is performed in Fig.\ \ref{fig:2}, where the
theoretical predictions have been normalised to the central measurements.
Theoretical uncertainties based on scale variations are shown in addition
as shaded areas. As expected, the LO uncertainties are considerably larger
than those at NLO and NLO+PS, which demonstrates that the agreement of the
LO central values is rather accidental and depends on the tuning. The NLO
JETPHOX predictions have a different shape in the second rapidity bin and
globally underestimate the data even within error bars. This is even more the case
for NNPDF2.3QED NLO PDFs (not shown, but known \cite{cacciari}). The results
with MMHT2014, CT14 and NNPDF3.0 PDFs are very similar \cite{Aaboud:2017cbm}.
While the central POWHEG NLO+PS predictions still underestimate the data
in the central rapidity bin, they generally agree best with the data and
always within scale uncertainties. Since our NLO calculation of direct
photon and purely partonic processes in POWHEG agrees with those in JETPHOX
\cite{Jezo:2016ypn}, this seems to indicate that the PS fragmentation in
PYTHIA 8 describes the data slightly better than the BFG II fragmentation
function.

\subsection{Photon-jet correlations in pp and pPb collisions with CMS}
\label{sec:3.2}

Compared to inclusive photon production, photon-jet correlations are more
sensitive to initial- and final-state QCD effects like PDFs, photon and jet
fragmentation, and their modifications in nuclear collisions. They have
therefore recently been measured by the CMS collaboration in pp and PbPb
collisions at a nucleon-nucleon centre-of-mass energy of $\sqrt{s_{NN}}=2.76$
TeV and in pPb collisions at a nucleon-nucleon centre-of-mass energy of
$\sqrt{s_{NN}}=5.02$ TeV \cite{CMS:2013oua}. During 2013 (pp), 2011 (PbPb), and 2013 (pPb),
integrated luminosities of 5.3 pb$^{-1}$, 150 $\mu$b$^{-1}$, and 30.4 nb$^{-1}$
were accumulated, respectively. The pp collisions are usually
considered to be the (vacuum) baseline with equal photon and jet transverse
momentum at LO, although this no longer holds at NLO and beyond (see below).
``Hot'' PbPb collisions are then studied to establish the creation of a QGP
and to determine its properties, in particular through jet quenching (parton energy loss),
while ``cold'' pPb collisions allow to check the absence of nuclear effects
in the final state and to quantify the modifications of the PDFs in the
initial state.

Similarly to the ATLAS analysis described above, CMS applied an isolation criterion
to the photons, albeit with a fixed cut on the {\rm total} energy $E^{\rm iso}$ at 5 GeV.
In addition, photons which matched a track from charged particles
within $|\eta^\gamma-\eta^{\rm Track}|<0.02$
and $|\phi^\gamma-\phi^{\rm Track}|<0.15$ were discarded in order to remove 
contamination from electrons, and the remaining photons had to satisfy $p_T^\gamma>40$
GeV and $|\eta^\gamma|<1.44$.
Jets were defined with an anti-$k_T$ cluster algorithm \cite{Cacciari:2008gp}
and a distance parameter $R=0.3$ and had to satisfy $p_T^{\rm jet}>30$ GeV and
$|\eta^{\rm jet}|<1.6$ to remain in the barrel calorimeter.

\begin{figure}[!t]
 \includegraphics[width=\linewidth]{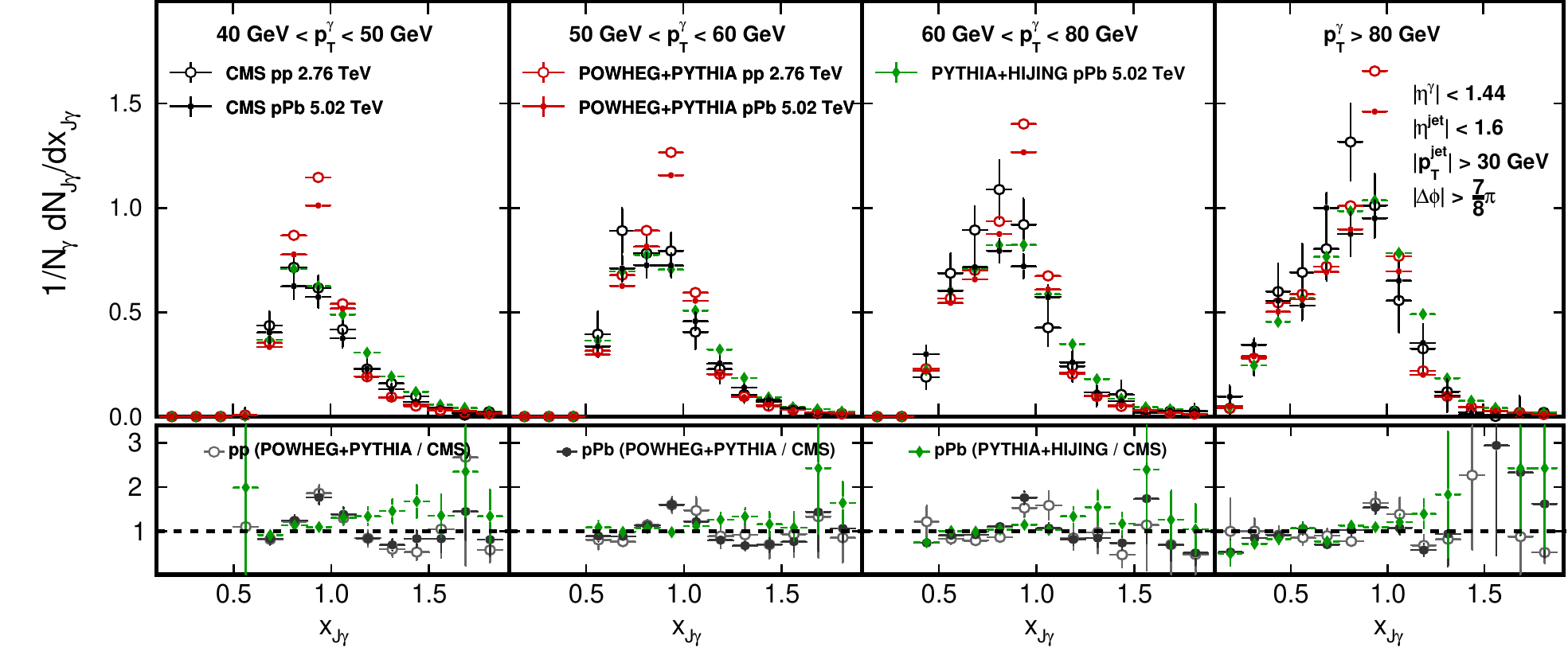}
 \caption{\label{fig:3}Distributions in the transverse momentum ratio of jets
 over photons in pp (open circles) and pPb (full circles) collisions at the LHC
 with a centre-of-mass energy per nucleon of
 $\sqrt{s_{NN}}=2.76$ and 5.02 TeV, respectively, in four different bins of
 photon transverse momentum. CMS 2013 data (black) are
 compared to predictions in LO with PYTHIA+HIJING (green) and NLO+PS
 with POWHEG+PYTHIA (red). The lower panels show the ratios of theory over data
 for pp (open circles) and pPb (full circles and green diamonds).}
\end{figure}

The distributions in Fig.\ \ref{fig:3} depend on the transverse momentum
ratio of jets over photons $x_{\rm J\gamma}=p_T^{\rm jet}/p_T^\gamma$, which
in LO would equal unity.
%++
 If the photon is associated with several jets, each of them contributes in the histograms,
 but the main contribution comes from photon plus one jet events. The reason is that
%--
%Here,
a cut on $\Delta\phi_{\rm J\gamma}>7\pi/8$ is applied
%++
 to each of the jets
%--
to suppress contributions from background and additional jets.
Since the pp (open circles) and pPb (full circles) data are consistent
with each other and the PYTHIA+HIJING
\cite{Wang:1991hta} Monte Carlo simulation, nuclear effects and centre-of-mass
energy dependence must be
subdominant. However, the jets become softer at larger photon transverse
momenta relative to the photon. Our NLO+PS calculations with POWHEG+PYTHIA describe the
data %very
well. They have a slightly sharper peak at somewhat larger
values of $x_{\rm J\gamma}$, since they have not been smeared with the
resolution of the CMS detector.

\begin{figure}[!h]
 \includegraphics[width=\linewidth]{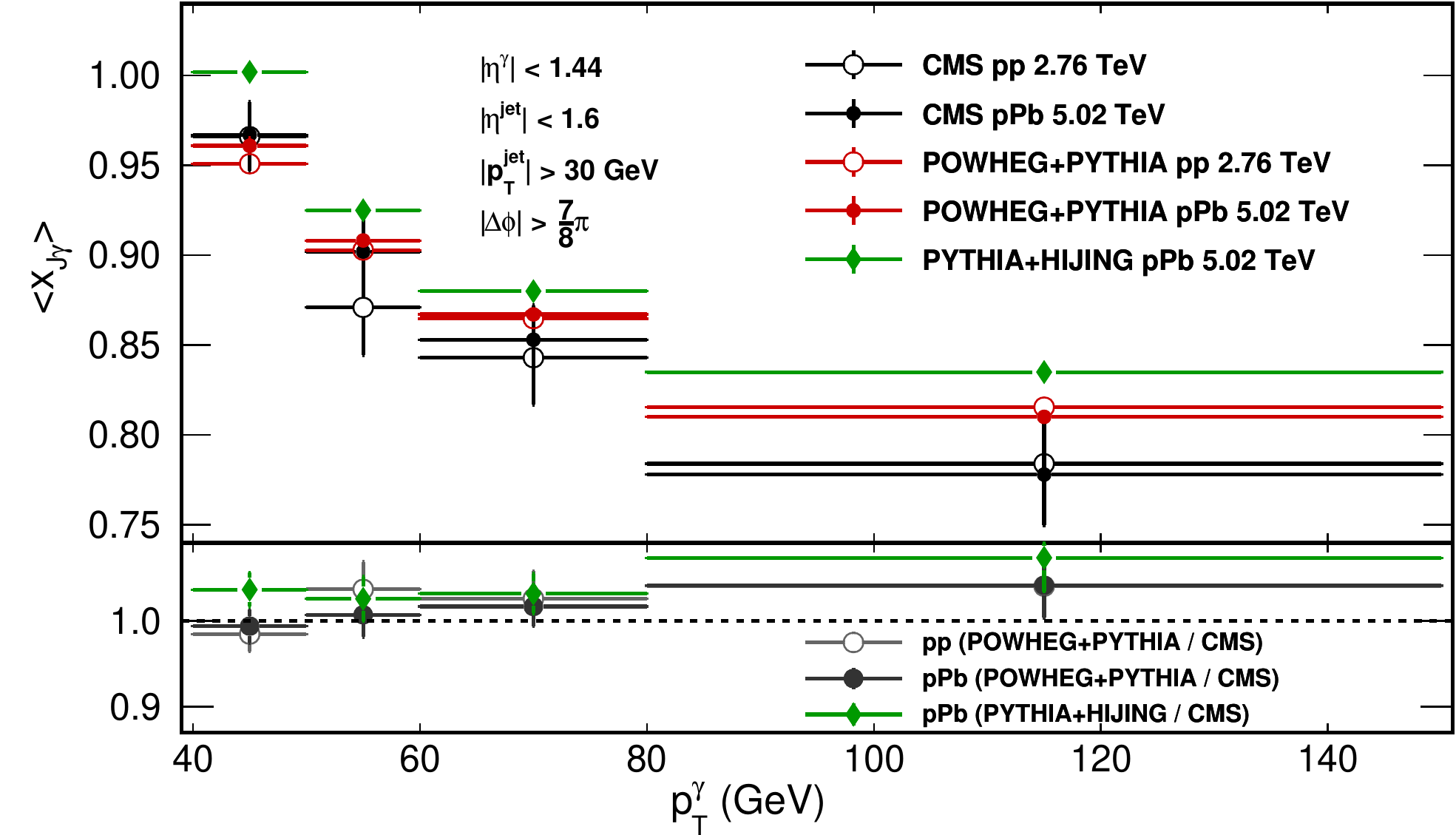}
 \caption{\label{fig:4}Average jet transverse momentum fraction as a
 function of photon transverse momentum. CMS 2013 data (black)
 are compared to predictions in LO with PYTHIA+HIJING (green) and NLO+PS
 with POWHEG+PYTHIA (red). The rightmost $p_T^\gamma$-bin refers to
 $p_T^\gamma>80$ GeV (no upper bound). The lower panel shows the
 ratio of theory over data for pp (open circles) and pPb (full circles
 and green diamonds), which coincide partially for the largest $p_T^\gamma$-bin.}
\end{figure}

The average values of $x_{\rm J\gamma}$ are plotted in Fig.\ \ref{fig:4} as
a function of the photon transverse momentum. Again, there are no
significant differences between the pp and pPb data, supporting the
interpretation that no QGP is created and rescattering effects in
the medium can thus not occur. This is in contrast to a more recent
measurement by CMS in PbPb collsions at $\sqrt{s_{NN}}=5.02$ TeV, which
is considerably lower than the pp baseline even after accounting
for different jet resolutions \cite{CMS:2016ynj}. At higher $p_T^\gamma$, several jets
can jointly balance the transverse momentum of the photon, e.g.\ in
a ``Mercedes star'' configuration, which leads to the observed falling
distribution in $\langle x_{\rm J\gamma}\rangle$.

\begin{figure}[!h]
 \includegraphics[width=\linewidth]{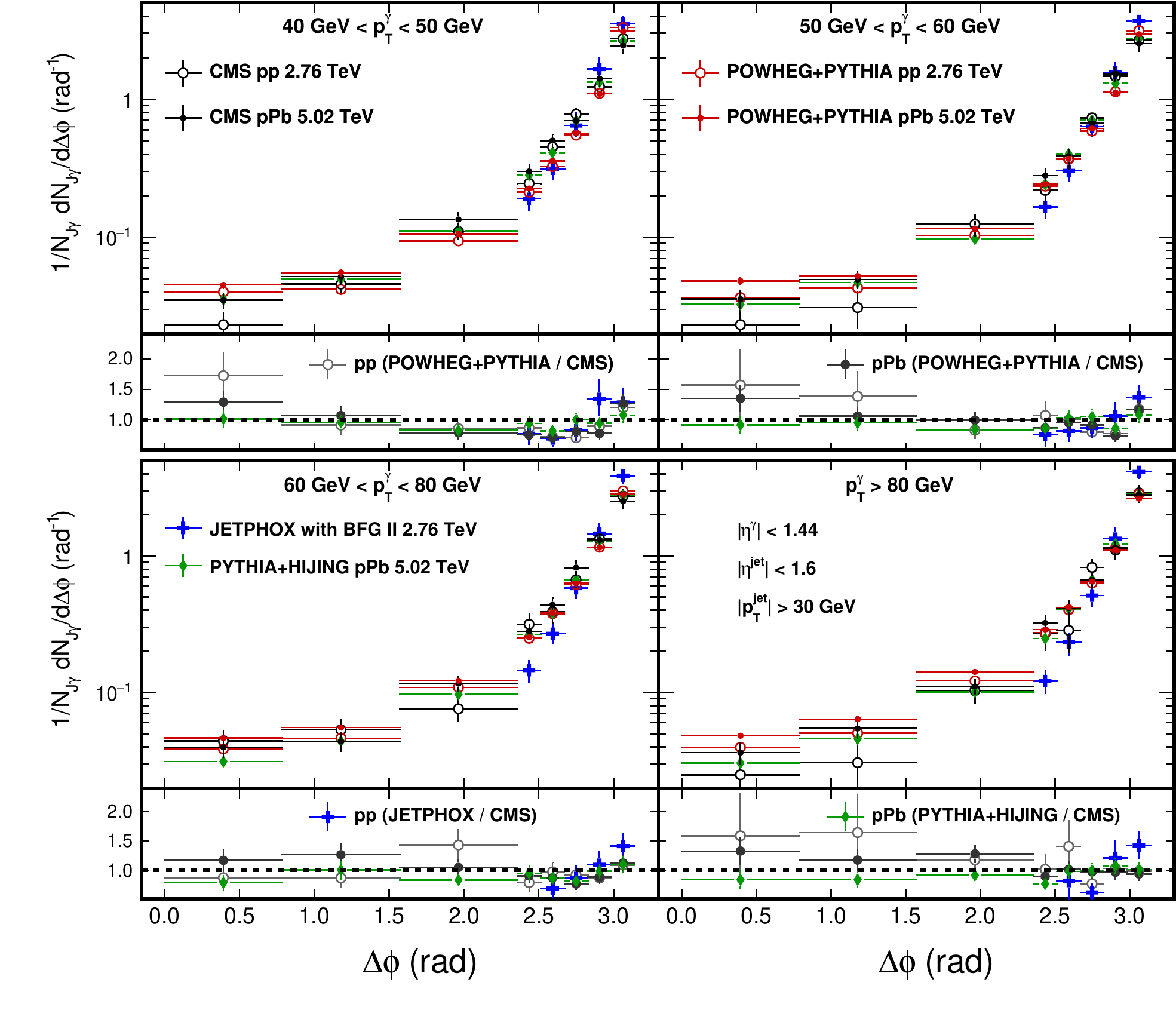}
 \caption{\label{fig:5}Relative azimuthal angle distributions of jets and
 photons in pp (open circles) and pPb (full circles) collisions at the LHC
 with a centre-of-mass energy per nucleon of
 $\sqrt{s_{NN}}=2.76$ and 5.02 TeV, respectively, in four different bins of
 photon transverse momentum. CMS 2013 data (black) are
 compared to predictions in LO with PYTHIA+HIJING (green), NLO with JETPHOX
 (blue), and NLO+PS with POWHEG+PYTHIA (red). The lower panels show the
 ratios of theory over data for pp (open circles and blue crosses) and pPb
 (full circles and green diamonds).}
\end{figure}

Possible medium effects of the back-to-back photon and recoiling
jet alignment can also be studied by comparing relative azimuthal
angle ($\Delta\phi_{J\gamma}$) distributions in proton and heavy-ion
collisions. Fig.\ \ref{fig:5} displays these distributions in pp
and pPb collisions in the same four bins of photon transverse
momentum used above. Again, no significant cold nuclear effects
or centre-of-mass
energy dependences are visible in the data. The LO predictions with
PYTHIA+HIJING agree well with the measurements, as do those at NLO+PS
with POWHEG+PYTHIA. Note that the near-side region at small $\Delta
\phi_{J\gamma}$ is quite sensitive to the experimental conditions on
the photon isolation. At larger $p_T^\gamma$, the distributions become
slightly flatter, which can again be traced to multi-jet configurations
(note that here no cut on $\Delta\phi_{\rm J\gamma}>7\pi/8$ is applied).
Azimuthal angle distributions can also be computed with JETPHOX, but
they do not describe the data below $\Delta\phi_{J\gamma}=2\pi/3$, as the
maximum number of jets is limited to two at NLO. 

\begin{figure}[!h]
 \includegraphics[width=\linewidth]{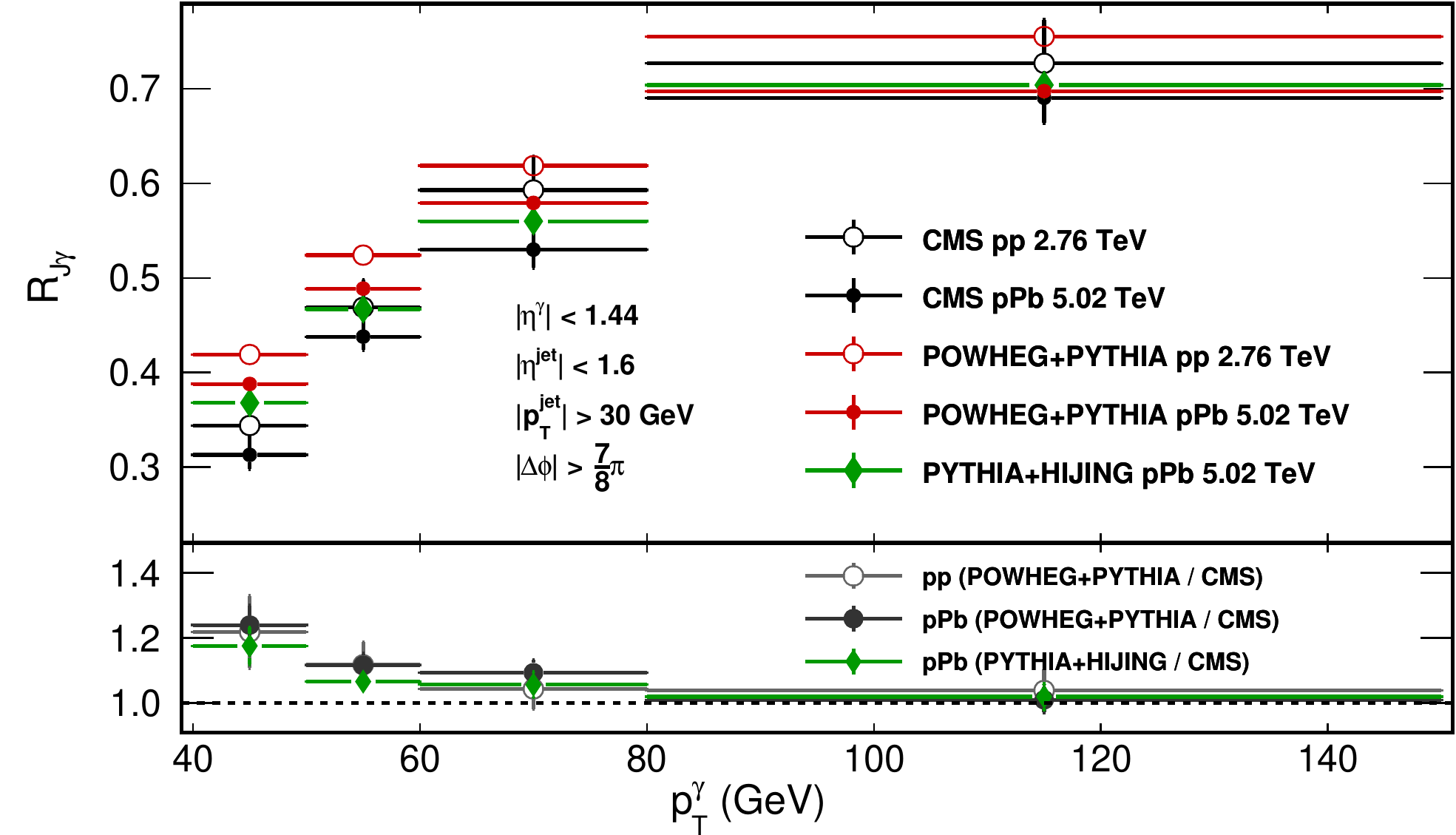}
 \caption{\label{fig:6}Fraction of photons associated to a jet
 with $p_T^{\rm jet}>30$ GeV as a function
 of photon transverse momentum in pp (open circles) and pPb (full circles)
 collisions at the LHC with a
 centre-of-mass energy per nucleon of $\sqrt{s_{NN}}=2.76$ and 5.02 TeV,
 respectively. CMS 2013 data (black) are compared to
 predictions in LO with PYTHIA+HIJING (green) and NLO+PS with
 POWHEG+PYTHIA (red). The lower panel shows the
 ratio of theory over data for pp (open circles) and pPb (full circles
 and green diamonds).}
\end{figure}

Not all photon events in the CMS measurement include a jet that passes
the cut on $p_T^{\rm jet}> 30$ GeV. The fraction of isolated photons that
are associated to a jet is therefore less than unity and shown in Fig.\
\ref{fig:6} as a function of the photon transverse momentum.
Here, again a cut on $\Delta\phi_{\rm J\gamma}>7\pi/8$
is applied to suppress contributions from background and additional jets.
Within the
experimental uncertainties, neither nuclear nor centre-of-mass energy
dependences are observed, although pPb collisions globally seem to have a
smaller $R_{J\gamma}$. %, possibly due to a larger number of softer (sometimes called ``mini'') jets.
%++
The pPb cross section is affected by nuclear shadowing (in particular
of the gluons) at $x_T^\gamma\geq2\times40$ GeV$/5020$ GeV $\geq 0.016$. This
not only reduces the inclusive photon cross section (which would not affect
$R_{J\gamma}$, since most photons recoil against jets), but fewer initial gluons
also lead to fewer quark jets, which would pass the cut on $p_T^{\rm jet}>30$
GeV more easily than gluon jets (which does affect $R_{J\gamma}$).
%--
This (statistically not significant) observation
is reproduced by our NLO POWHEG+PYTHIA calculations, but they slightly
overestimate both the pp and pPb data. %, since we have not included multiple scatterings or soft underlying events. As expected, the LO PYTHIA+HIJING simulations, which include these effects, describe the data better.
Note that the ATLAS collaboration has recently also presented a
measurement of isolated-photon and jet production, but only in pp
collisions at $\sqrt{s}=13$ TeV \cite{ATLAS:2017gxj}. 

%%%%%%%%%%%%%% Begin Section 4 %%%%%%%%%%%%%%%%%%%%%%%%%%%%%%%%%%%%%%%%%
\section{Predictions for future LHC analyses and sensitivity to nuclear PDFs}
\label{sec:4}

In this section we make NLO+PS predictions for future LHC analyses and
compare them to those at LO+PS and NLO. First, we study the impact of
higher-order corrections on the ratio of inclusive over decay photons.
This quantity has been shown by ALICE to be of crucial importance in
the determination of the effective temperature of the QGP. Second, we
study the sensitivity of photon+jet measurements in pPb collisions with
ALICE, ATLAS, CMS and LHCb to nuclear PDFs. In particular, we show that
distributions in the observed momentum fraction of the parton in the lead
ion $x_{\rm Pb}^{\rm obs}$ are sensitive to nuclear effects like shadowing
at low $x$. Related variables have previously been used successfully for
proton and photon PDF determinations at the HERA collider at DESY.

\subsection{Ratio of inclusive over decay photons with ALICE}
\label{sec:4.1}

Inclusive photons represent an essential probe of the medium produced in central
heavy-ion collisions, since the thermal contribution to the exponentially falling
spectrum at low transverse momenta can be related to an effective temperature of
the QGP. In contrast, the hard perturbative contributions, e.g.\ from QCD Compton
and photon-gluon fusion processes, lead to a power-law behaviour in $p_T^\gamma$.
The corresponding measurements in pp collisions serve as a (vacuum) baseline and
have been performed with 2010 and 2011 data by the ALICE collaboration at
centre-of-mass energies of 0.9, 2.76 and 7 TeV \cite{ALICE:2014rma}. To remain
sensitive to the thermal radiation, the photons must, however, not be isolated,
leading to the complication that the spectra are dominated by contributions from
neutral pion, $\eta$-, and $\omega$-meson decays.

Decay photon contributions were subtracted only in a subsequent ALICE measurement
of PbPb collisions at $\sqrt{s_{NN}}=2.76$ TeV \cite{Adam:2015lda}. This measurement
was reanalysed in NLO and including theoretical uncertainties in Ref.\ \cite{Klasen:2013mga}.
The crucial quantity in the subtraction procedure was the $p_T^\gamma$-dependent
double ratio
\beq
 R_\gamma={\gamma_{\rm incl.}\over\pi^0_{\rm param.}}/{\gamma_{\rm decay}\over\pi^0_{\rm param.}},
\eeq
where $\gamma_{\rm incl.}$ is the measured inclusive photon spectrum, 
$\pi^0_{\rm param.}$ is a parameterisation of the measured $\pi^0$ spectrum,
and $\gamma_{\rm decay}$ is the calculated decay photon spectrum. It has the
advantage that some of the largest systematic uncertainties cancel.
The quantity $R_\gamma$ allowed to measure prompt (i.e.\ non-decay) photons with
0.9 GeV $<p_T^\gamma<14$ GeV and $|\eta^\gamma|<0.9$
in PbPb collisions through $\gamma_{\rm prompt}=(1-1/R_\gamma)\gamma_{\rm incl.}$ for
different centralities and to extract an effective temperature of the QGP in
the most central collisions.
A similar measurement is currently in progress for PbPb collisions at
$\sqrt{s_{NN}}=5.02$ TeV.

\begin{figure}[!t]
 \includegraphics[width=\linewidth]{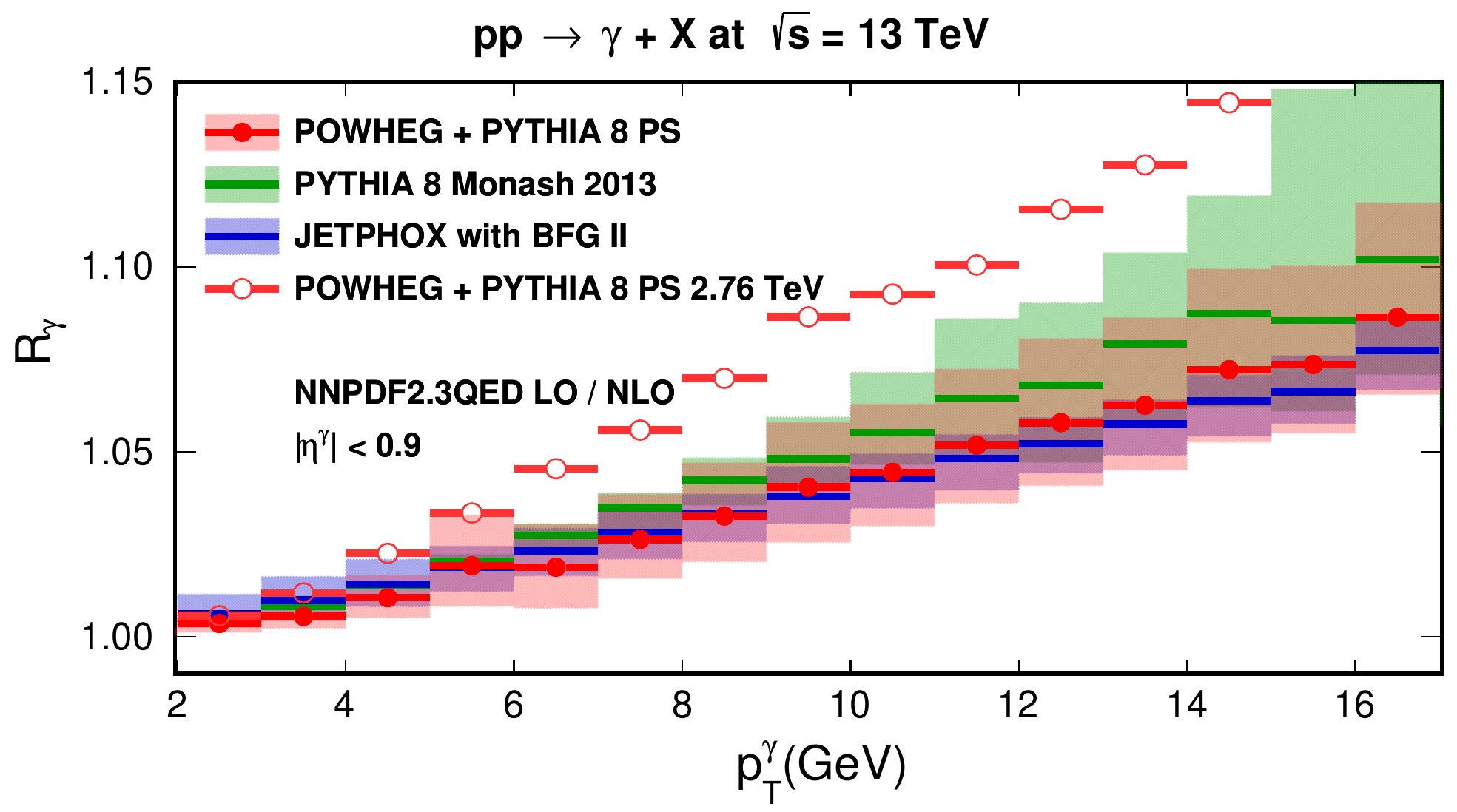}
 \caption{\label{fig:7}Ratio of inclusive over decay photon production in pp
 collisions at the LHC with a centre-of-mass energy of $\sqrt{s}=13$ TeV.
 LO predictions with PYTHIA (green) are compared to those at NLO with JETPHOX
 (blue) and NLO+PS with POWHEG+PYTHIA (red, full circles). The latter are
 also shown at $\sqrt{s} = 2.76$ TeV (red, open circles).}
\end{figure}

Since $R_\gamma$ relied on LO Monte Carlo simulations of the decay photon
spectrum with PYTHIA, it is important to quantify the impact of higher-order
effects on this observable. In Fig.\ \ref{fig:7} we show simulations of the
(simplified) theoretical ratio
\beq
 R_\gamma={\gamma_{\rm incl.}\over\gamma_{\rm decay}}=1+{\gamma_{\rm prompt}\over\gamma_{\rm decay}}
\eeq
for the current pp run at $\sqrt{s}=13$ TeV. This quantity compared to data
will help to constrain additional sources of low-$p_T^\gamma$ photons at
the highest LHC energy, allowing to validate the thermal origin of photons
measured in PbPb collisions. In Fig.\ \ref{fig:7}, the prompt photon spectrum
is calculated in LO with PYTHIA (green), NLO with JETPHOX and BFG II
photon fragmentation functions (blue), and NLO+PS with POWHEG+PYTHIA (red,
full circles),
respectively. In contrast, the decay photon spectrum is always calculated
in LO with the Monash 2013 tune of PYTHIA 8, which is known to describe
the pion data very well, and does not contribute to the theoretical
uncertainty. Once measured, the decay contribution should be replaced by the
experimental values.
%Note that both JETPHOX and POWHEG not only allow for calculations of inclusive photon, but also inclusive hadron production.
For better comparability, NNPDF2.3QED PDFs
are used in all calculations, since initial-state PDF effects are
expected to cancel to a large extent. As expected, the scale uncertainties
(shaded bands) are considerably larger at LO than at NLO and NLO+PS. The
smaller scale uncertainty in NLO compared to NLO+PS is due to an additional
cancellation of factorisation scale dependence in the fragmentation
contribution. For $p_T^\gamma>7$ GeV, the central values of $R_\gamma$ at LO
are larger than those at NLO and NLO+PS and even fall outside the error
band of the former. In contrast, the central values at NLO and NLO+PS
agree very well and should therefore be employed in future experimental
measurements.

For illustration we also show in Fig.\ \ref{fig:7} the POWHEG+PYTHIA
results for pp collisions at $\sqrt{s} = 2.76$ TeV (red, open circles).
The direct photon signal is much more significant at lower $\sqrt{s}$
mainly due to the steeper spectra and larger probed $x$-values. Its
contribution relative to the one by decay photons $(R_\gamma-1)$ changes
for typical LHC energies from 13 TeV to 2.76 TeV by about a factor of two.
A similar energy dependence is expected for the direct photon signal
in PbPb collisions, since neither $T_{AA}$ nor the suppression of the
hadronic decay background vary strongly with $\sqrt{s}$. The suppression
of the decay background due to partonic energy loss in PbPb collisions
leads in general to $R_\gamma$ values larger by factors of five to seven compared
to pp collisions.

\subsection{Photon-jet production at high $p_T^\gamma$ in pPb collisions with ATLAS or CMS}
\label{sec:4.2}

In the remaining three subsections, we make predictions for associated
photon and jet production in proton-lead collisions at a centre-of-mass
energy per nucleon of $\sqrt{s_{NN}} = 5.02$ TeV. Our focus will be on
the potential of the four major LHC experiments ALICE, ATLAS, CMS and
LHCb to improve on current constraints of nuclear PDFs with measurements
of this reaction. In particular, we propose to analyse the experimental
data in terms of distributions in the observed parton momentum fraction
of the lead ion, defined as
\beq
 x_{\rm Pb}^{\rm obs} := { p_T^\gamma e^{-\eta^\gamma} + p_T^{\rm jet} e^{-\eta^{\rm jet}}
 \over
 2 E_{\rm Pb}},
\eeq
so that regions of nuclear shadowing and antishadowing, EMC effect and
Fermi motion can directly be located in increasing regions of $x$.
Related variables have been used successfully at DESY HERA for determinations
of proton and photon PDFs \cite{Klasen:2002xb}.

We base our predictions on the current status of nuclear PDF determinations
from deep-inelastic scattering (DIS) and Drell-Yan (DY) measurements and
use for our central estimates the recent fit nCTEQ15(np) \cite{Kovarik:2015cma}.
It is based on proton PDFs similar to CTEQ6.1  \cite{Stump:2003yu}, but
with minimal influence from nuclear data. For comparison, we will also show
nCTEQ15 estimates with additional constraints from inclusive pion data taken
by the PHENIX and STAR experiments at RHIC as well as estimates with
the latest EPPS16 nuclear PDF fit \cite{Eskola:2016oht} that already includes
LHC data from ATLAS and CMS on electroweak gauge boson and dijet production
in pPb collisions. The nCTEQ and EPPS16 fits come with 32 and 40 error PDFs,
respectively, which we will show as shaded bands. We will, however, not
vary the underyling proton PDF, assuming that it can be sufficiently
constrained from pp collisions by the same experiments. Instead, we will
show for comparison the pp baseline at $\sqrt{s}=2.76$ TeV with theoretical
uncertainties obtained by scale variations of relative factors of two, but
not four.

\begin{figure}[!h]
 \includegraphics[width=\linewidth]{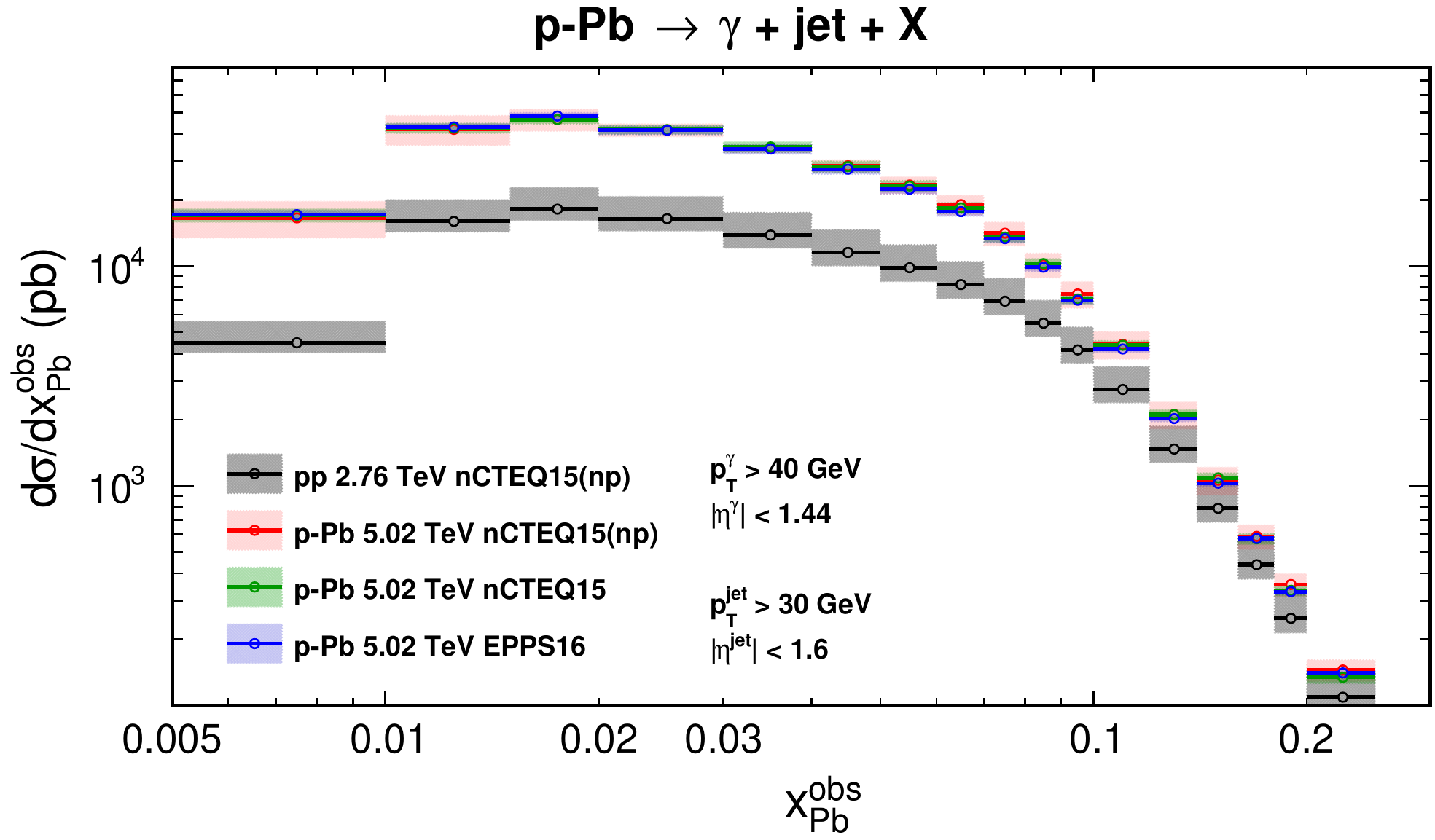}
 \caption{\label{fig:8}Nuclear PDF sensitivity of central high-$p_T^\gamma$
 photon-jet measurements with ATLAS or CMS in pPb collisions at the LHC
 with a nucleon-nucleon centre-of-mass energy of $\sqrt{s_{NN}}=5.02$ TeV.
 Predictions with nCTEQ15(np) PDFs (red) are compared to those with
 nCTEQ15 including pion data from RHIC (green) and EPSS16 including weak
 boson and jet data from LHC (blue). The nCTEQ15 pp baseline is also shown
 (grey).}
\end{figure}

In this subsection, we concentrate on the ATLAS and CMS experimental acceptances.
In particular, we require the photons to be isolated with $E^{\rm iso}<5$
GeV inside a cone of size $\Delta R=0.4$, to have relatively high transverse
momenta of $p_T^\gamma>40$ GeV, and to remain in the CMS barrel electromagnetic
calorimeter with $|\eta^\gamma| < 1.44$. Jets, defined by the anti-$k_T$
cluster algorithm with distance parameter $R=0.3$, have to satisfy
$p_T^{\rm jet}>30$ GeV and remain in the CMS hadronic barrel calorimeter with
$|\eta^{\rm jet}| < 1.6$. These cuts correspond to those used in Sec.\ 
\ref{sec:3.2} \cite{CMS:2013oua}.  The ATLAS acceptances with
$|\eta^\gamma| < 1.475$ and $|\eta^{\rm jet}| < 1.7$ are very similar
\cite{ATLAS:2017gxj}.
The results of this proposed analysis are shown in Fig.\ \ref{fig:8}.
The observable region of the parton momentum fraction in the lead ion
ranges from 0.005 to 0.25. The cross sections at $\sqrt{s}=2.76$ TeV
are lower than those at 5.02 TeV by up to a factor of five at low $x$.
At NLO+PS, the scale uncertainties
in pp collisions (grey) are still sizeable, so that they should be
eliminated -- like the proton PDF uncertainties (not shown) -- by a
corresponding baseline measurement or by taking ratios of nuclear and
bare proton cross sections. The shadowing region below $x\sim0.02$ is not
well constrained by DIS and DY data (red) and exhibits uncertainties
of up to a factor of two. The situation improves considerably after
inclusion of hadron collider data from RHIC (green) or LHC (blue),
which can also be expected from the proposed inclusive photon measurements.
Similar improvements are observed in the antishadowing region above
$x\sim0.05$. In the intermediate region, nuclear effects are
known to be small.

\begin{figure}[!h]
 \includegraphics[width=\linewidth]{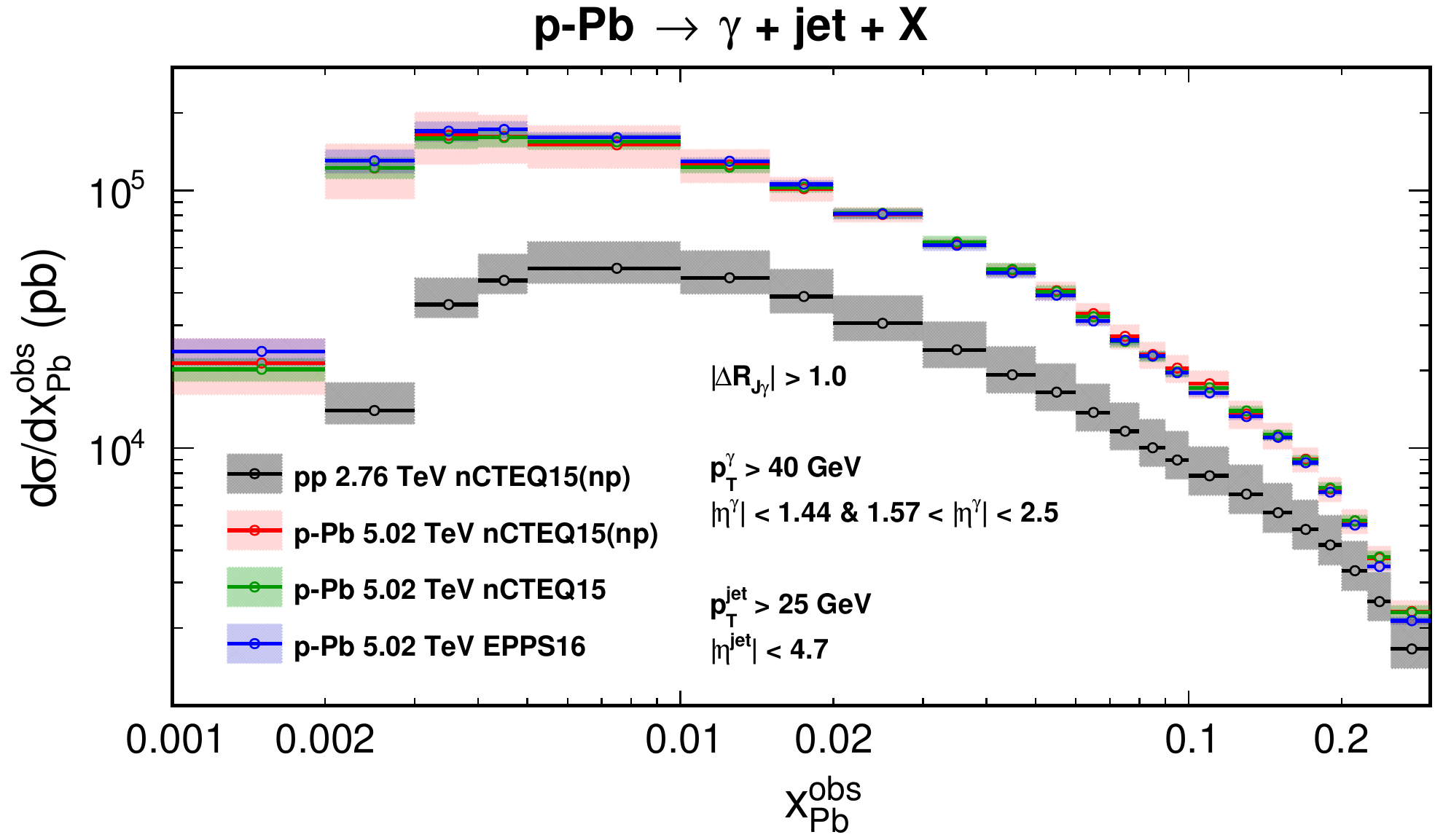}
 \caption{\label{fig:12}Same as Fig.\ \ref{fig:8}, but with larger
 acceptances in photon and jet rapidity, using endcap calorimeters,
 and lower threshold on the jet transverse momentum.}
\end{figure}

Increasing the photon acceptance to the CMS electromagnetic endcap
calorimeter $|\eta^\gamma| < 2.5$, but excluding the transition
region of $1.44<|\eta|<1.57$, and the jet acceptance to
$p_T^{\rm jet}>25$ GeV and $|\eta^{\rm jet}| < 4.7$ as in Ref.\
\cite{CMS:2015hha} leads to an extension of the observable $x$-range
in the lead ion down to 0.001, as shown in Fig.\ \ref{fig:12}.
Here, jets are in addition required to have a distance $\Delta R_{J\gamma}
>1.0$ from the photon. As expected, the uncertainties from nuclear
effects increase below $x$-values of 0.005. This holds now not
only for the fit of DIS and DY data only (red), but also for the
fits that include RHIC (green) or LHC (blue) data, demonstrating
the potential of inclusive photon and jet measurements to constrain
the nuclear PDFs even further. The central fits of nCTEQ15(np),
nCTEQ15 and EPPS16 are now also clearly distinguishable, the
RHIC data favouring more, the LHC data less nuclear shadowing
than the DIS and DY data alone.

\subsection{Photon-jet production at low $p_T^\gamma$ in pPb collisions with ALICE}

As discussed in Sec.\ \ref{sec:4.1}, the region of low transverse momenta
gives access to the thermal photon spectrum produced in PbPb collisions.
In addition, it should allow to probe even deeper into the low-$x$
shadowing region in pPb collisions. The low-$p_T$ region is well
instrumented in the ALICE detector, so that in this subsection we
apply the corresponding cuts on photons, isolated in a cone of size
$R < 0.4$ with $E_T^{\rm iso}<2$ GeV \cite{germain}, of $p_T^\gamma > 1$ GeV
and $|\eta^\gamma| < 0.9$ \cite{Adam:2015lda} and on jets, defined
by the anti-$k_T$ algorithm with $R=0.4$, of $p_T^{\rm jet} \in [20;120]$
GeV and $|\eta^{\rm jet}| < 0.5$ \cite{Adam:2015hoa}.

\begin{figure}[!h]
 \includegraphics[width=\linewidth]{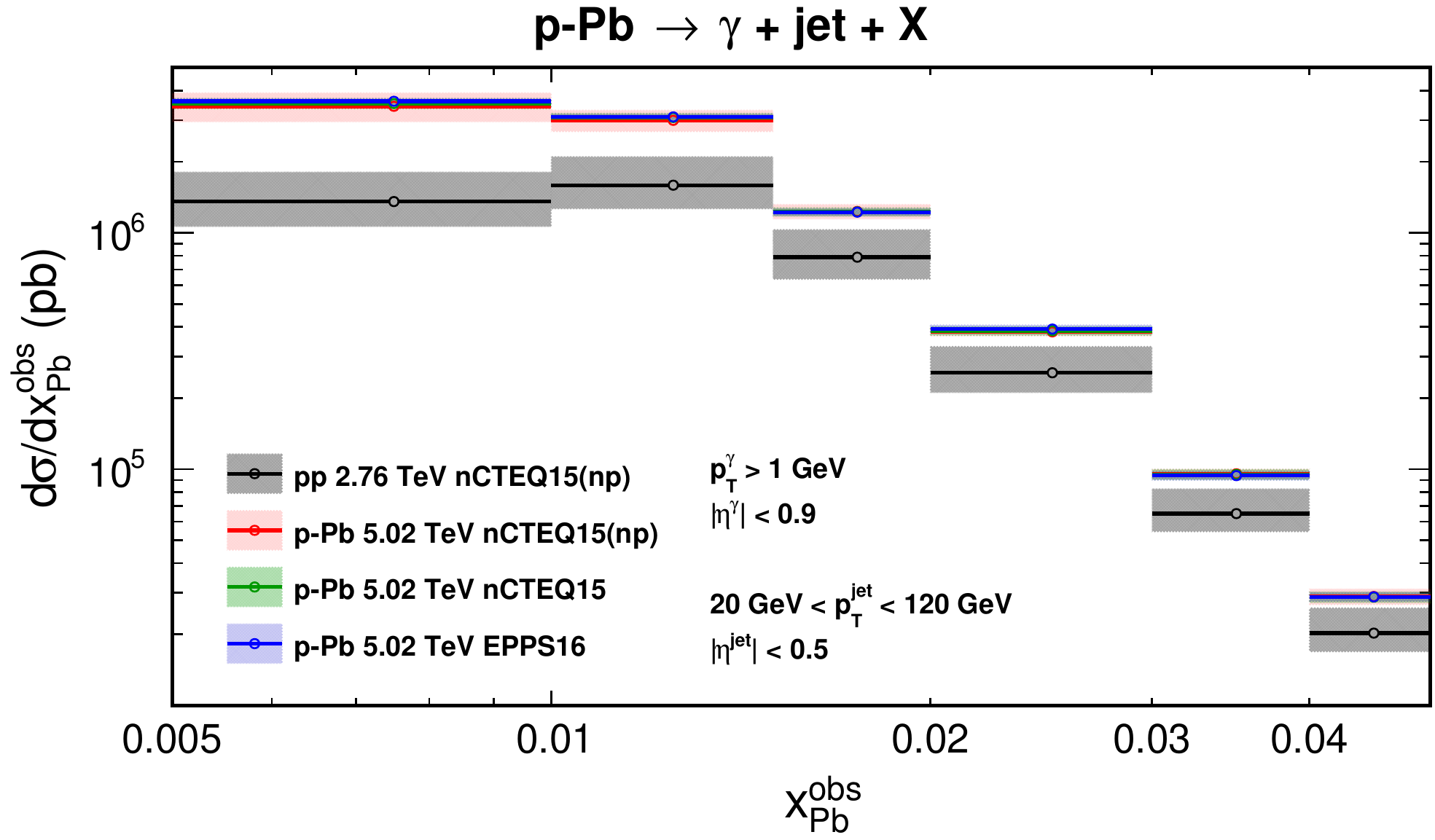}
 \caption{\label{fig:9}Nuclear PDF sensitivity of central low-$p_T^\gamma$
 photon-jet measurements with ALICE in pPb collisions at the LHC with a
 nucleon-nucleon centre-of-mass energy of $\sqrt{s_{NN}}=5.02$ TeV.
 Predictions with nCTEQ15(np) PDFs (red) are compared to those with
 nCTEQ15 including pion data from RHIC (green) and EPSS16 including weak
 boson and jet data from LHC (blue). The nCTEQ15 pp baseline is also shown
 (grey).}
\end{figure}

Fig.\ \ref{fig:9} shows that this ALICE acceptance does not lead to
lower observable values of $x$, but only to a similar $x$-range
as described in the previous subsection, Sec.\ \ref{sec:4.2}, for
central high-$p_T^\gamma$ analyses with ATLAS and CMS, contrary to
what one might expect for lower $p_T^\gamma$. The reason
is that the ALICE acceptance in rapidity is considerably more
central than the one of ATLAS or CMS. It would therefore be
benefitial if one could also use the ALICE Photon Multiplicity
Detector (PMD) situated at forward rapidities of $2.3<\eta^\gamma<3.9$
\cite{ALICE:2014rma}.

\subsection{Forward photon-jet production in p-Pb and Pb-p collisions with LHCb}

Although the main focus of the fourth LHC experiment LHCb is on
precisision measurements of rare $B$ decays, the very forward
instrumentation of the LHCb detector should also permit interesting
measurements of other observables, that are complementary to the
three other LHC experiments. Indeed, inclusive photons have been
measured in radiative $B^0_s$ decays, even with polarisation, with
$p_T^\gamma > 3$ GeV and $\eta^\gamma \in [2;5]$ \cite{Aaij:2016ofv}.
For comparability with the previous subsection, we employ in
addition the same photon isolation criterion as used by the ALICE
collaboration \cite{germain}. Jets, defined by the anti-$k_T$ algorithm with
$R=0.5$, have been measured by LHCb (in association with forward
electroweak bosons) in the region $p_T^{\rm jet} \in [20;100]$ GeV
and $\eta^{\rm jet} \in [2.2;4.2]$ \cite{AbellanBeteta:2016ugk}.

\begin{figure}[!h]
 \includegraphics[width=\linewidth]{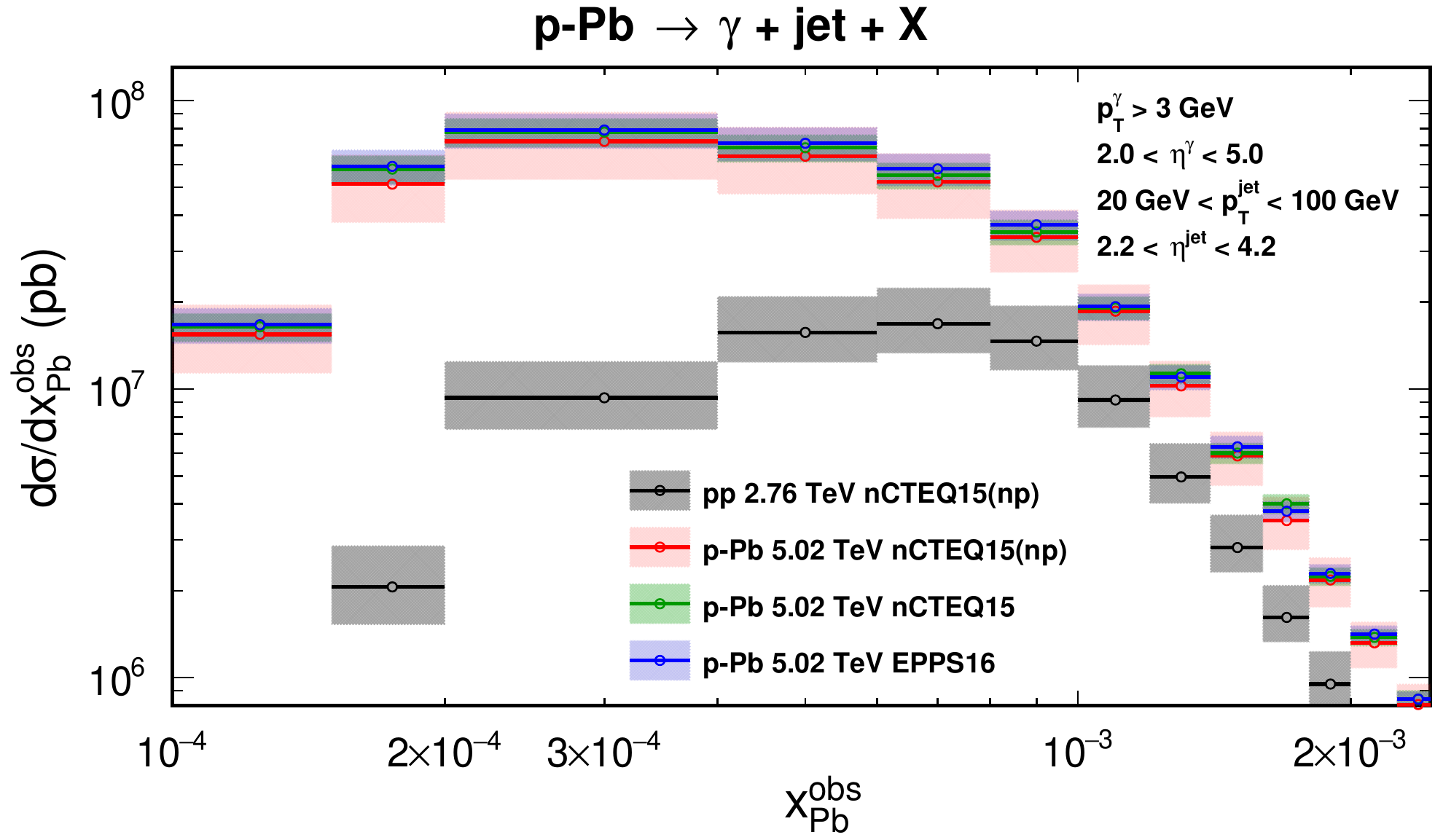}
 \caption{\label{fig:10}Nuclear PDF sensitivity of forward low-$p_T^\gamma$
 photon-jet measurements with LHCb in p-Pb collisions at the LHC with a
 nucleon-nucleon centre-of-mass energy of $\sqrt{s_{NN}}=5.02$ TeV.
 Predictions with nCTEQ15(np) PDFs (red) are compared to those with
 nCTEQ15 including pion data from RHIC (green) and EPSS16 including weak
 boson and jet data from LHC (blue). The nCTEQ15 pp baseline is also shown
 (grey).}
\end{figure}

We have employed these acceptances to p-Pb collisions
with the proton beam going in the positive $z$-axis direction. The lead
ion is therefore always probed in the very small-$x$ region between
$10^{-4}$ and $2\times10^{-3}$, as one can see in  Fig.\ \ref{fig:10}.
At $x\sim10^{-4}$, the theoretical nuclear PDF uncertainty amounts to
almost a factor of two for nCTEQ15(np) and 50\% for nCTEQ15 or EPPS16.
Interestingly, both RHIC and LHC data now push the central predictions
above those from DIS and DY data, i.e.\ towards less nuclear shadowing.

\begin{figure}[!h]
 \includegraphics[width=\linewidth]{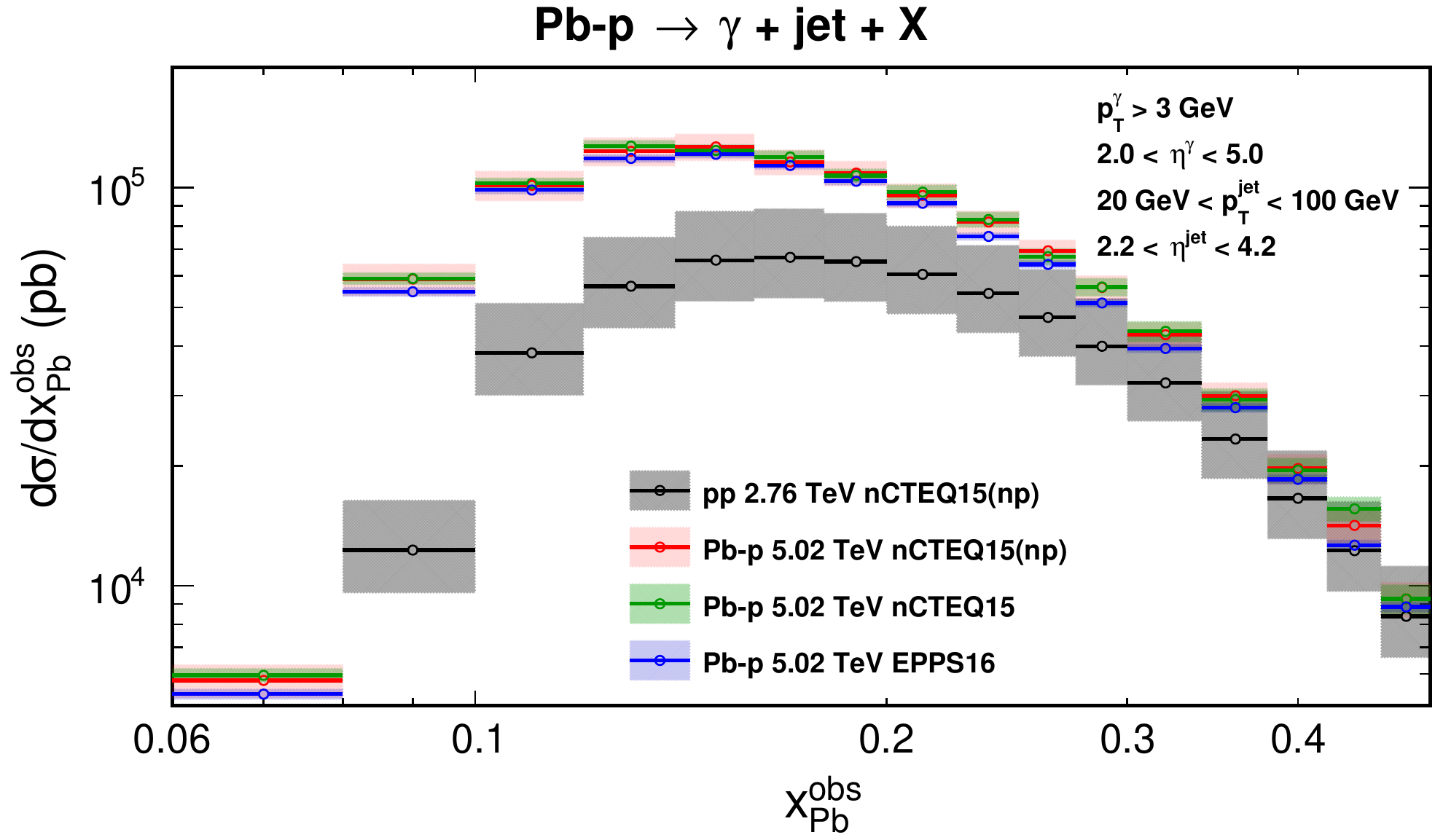}
 \caption{\label{fig:13}Same as Fig.\ \ref{fig:10}, but for Pb-p collisions.}
\end{figure}

When the lead ion is going in the positive $z$-axis direction, it is mostly
probed at large $x$, i.e.\ in the region of the EMC effect, as one can see
in Fig.\ \ref{fig:13}. This region is fairly well constrained
by DIS and DY data alone, since the data from the EMC DIS experiment
were heavily used in the nuclear PDF fits. So at large $x$, the
uncertainty bands overlap quite
well and shrink only moderately from nCTEQ15(np) to nCTEQ and EPPS16.
Consequently, measurements in this ``control'' region could be used to
fix the normalisation, i.e.\ the choices of scales or proton PDFs.

%%%%%%%%%%%%%% Begin Section 5 %%%%%%%%%%%%%%%%%%%%%%%%%%%%%%%%%%%%%%%%%
\section{Conclusions and outlook}
\label{sec:5}

Prompt photons have many important applications in high-energy collisions.
They range from determinations of the strong coupling constant, proton and
nuclear PDFs to
those of the properties of the QGP. Our recent calculation of prompt photon
and photon+jet production at NLO+PS with POWHEG+PYTHIA combines the
reliability of NLO calculations with detailed information on the final
state as required for realistic phenomenological analyses.

In this paper, we have performed a first comparison of these calculations
with LHC data, in particular of inclusive photon production as measured by
ATLAS and of photons produced in association with jets as measured by CMS
in pp and pPb collisions at various centre-of-mass energies of the LHC. For
inclusive photons, we found very good agreement and, as expected, considerably
reduced scale uncertainties compared to LO+PS predictions with PYTHIA. For
associated photon+jet production, the transverse momentum balance and azimuthal
correlations could be described for the first time correctly at NLO.

For the extraction of the thermal photon spectrum in PbPb collisions, the
ratio of inclusive over decay photons $R_\gamma$ represents an important
quantity that allows to eliminate the contribution from decay photons.
We predicted this quantity at NLO and NLO+PS, i.e.\ more reliably than it
was done previously by the ALICE collaboration, and thus reduced the
theoretical uncertainty considerably. The implementation of medium
effects in heavy-ion collisions is left for future work.

We then made predictions for pPb collisions for ongoing analyses at
$\sqrt{s_{NN}}=5.02$ TeV in all four major LHC experiments. In particular,
ALICE allows to measure also low transverse momenta, whereas LHCb has very
good forward instrumentation.  We studied the potential of these analyses
for future determinations of nuclear PDFs in distributions in the observed
parton momentum fraction in the lead ion $x_{\rm Pb}^{\rm obs}$. We found that
LHC photon data could have an important impact on the determination of
nuclear effects such as shadowing at low $x$.

%++
ATLAS has also measured isolated prompt photon production in PbPb collisions
at a nucleon-nucleon centre-of-mass energy of
$\sqrt{s_{NN}}=2.76$ TeV \cite{Aad:2015lcb}. After scaling the data by the
mean nuclear thickness $\langle T_{AA}\rangle$, they agree with NLO JETPHOX
predictions in all rapidity and centrality regions within statistical and
systematic uncertainties.
This supports the interpretation that the strong suppression observed for
the production of jets and hadrons in PbPb collisions compared to the
scaled pp measurement is due to a strong partonic energy loss in the medium.  
% The authors conclude that this supports the
%interpretation of a centrality-dependent jet quenching in the hot medium.
We expect this conclusion to hold also for our similar NLO+PS POWHEG+PYTHIA
results. A consistent matching of our NLO calculations with a medium-modified
PS is, however, beyond the scope of the present work.
%--

%%%%%%%%%%%%%% Begin Acknowledgment %%%%%%%%%%%%%%%%%%%%%%%%%%%%%%%%%%%%
\section*{Acknowledgments}

We thank M.\ Cacciari, D.\ d'Enterria and Yen-Jie Lee for useful discussions.
This work has been supported by the BMBF under contract 05H15PMCCA and by the  DFG
through the Research Training Network 2149 ``Strong and weak interactions -
from hadrons to dark matter''.

%%%%%%%%%%%%%% Begin Bibliography %%%%%%%%%%%%%%%%%%%%%%%%%%%%%%%%%%%%%%
\bibliographystyle{JHEP}

\bibliography{paper}

\begin{thebibliography}{}

%\cite{Albino:2002ck}
\bibitem{Albino:2002ck}
  S.~Albino, M.~Klasen and S.~S\"oldner-Rembold,
  %``Strong coupling constant from the photon structure function,''
  Phys.\ Rev.\ Lett.\  {\bf 89} (2002) 122004.
%  doi:10.1103/PhysRevLett.89.122004
%  [hep-ph/0205069].
  %%CITATION = doi:10.1103/PhysRevLett.89.122004;%%
  %29 citations counted in INSPIRE as of 05 Sep 2017

%\cite{Klasen:2002xb}
\bibitem{Klasen:2002xb}
  M.~Klasen,
  %``Theory of hard photoproduction,''
  Rev.\ Mod.\ Phys.\  {\bf 74} (2002) 1221.
%  doi:10.1103/RevModPhys.74.1221
%  [hep-ph/0206169].
  %%CITATION = doi:10.1103/RevModPhys.74.1221;%%
  %95 citations counted in INSPIRE as of 05 Sep 2017

%\cite{Stavreva:2010mw}
\bibitem{Stavreva:2010mw}
  T.~Stavreva, I.~Schienbein, F.~Arleo, K.~Kovarik, F.~Olness, J.~Y.~Yu and J.~F.~Owens,
  %``Probing gluon and heavy-quark nuclear PDFs with gamma + Q production in pA collisions,''
  JHEP {\bf 1101} (2011) 152;
%  doi:10.1007/JHEP01(2011)152
%  [arXiv:1012.1178 [hep-ph]].
  %%CITATION = doi:10.1007/JHEP01(2011)152;%%
  %43 citations counted in INSPIRE as of 05 Sep 2017
%
%\cite{Brandt:2014vva}
%\bibitem{Brandt:2014vva}
  M.~Brandt, M.~Klasen and F.~K\"onig,
  %``Nuclear parton density modifications from low-mass lepton pair production at the LHC,''
  Nucl.\ Phys.\ A {\bf 927} (2014) 78;
%  doi:10.1016/j.nuclphysa.2014.03.024
%  [arXiv:1401.6817 [hep-ph]].
  %%CITATION = doi:10.1016/j.nuclphysa.2014.03.024;%%
  %8 citations counted in INSPIRE as of 05 Sep 2017
%
%\cite{Klasen:2017kwb}
%\bibitem{Klasen:2017kwb}
  M.~Klasen, K.~Kovarik and J.~Potthoff,
  %``Nuclear parton density functions from jet production in DIS at an EIC,''
  Phys.\ Rev.\ D {\bf 95} (2017) 094013.
%  doi:10.1103/PhysRevD.95.094013
%  [arXiv:1703.02864 [hep-ph]].
  %%CITATION = doi:10.1103/PhysRevD.95.094013;%%
  %4 citations counted in INSPIRE as of 05 Sep 2017

%\cite{Arnold:2001ms}
\bibitem{Arnold:2001ms}
  P.~B.~Arnold, G.~D.~Moore and L.~G.~Yaffe,
  %``Photon emission from quark gluon plasma: Complete leading order results,''
  JHEP {\bf 0112} (2001) 009.
%  doi:10.1088/1126-6708/2001/12/009
%  [hep-ph/0111107].
  %%CITATION = doi:10.1088/1126-6708/2001/12/009;%%
  %369 citations counted in INSPIRE as of 05 Sep 2017

%\cite{Klasen:2013mga}
\bibitem{Klasen:2013mga}
  M.~Klasen, C.~Klein-B\"osing, F.~K\"onig and J.~P.~Wessels,
  %``How robust is a thermal photon interpretation of the ALICE low-p_T data?,''
  JHEP {\bf 1310} (2013) 119.
%  doi:10.1007/JHEP10(2013)119
%  [arXiv:1307.7034 [hep-ph]].
  %%CITATION = doi:10.1007/JHEP10(2013)119;%%
  %24 citations counted in INSPIRE as of 31 Aug 2017

%\cite{Wang:1996yh}
\bibitem{Wang:1996yh}
  X.~N.~Wang, Z.~Huang and I.~Sarcevic,
  %``Jet quenching in the opposite direction of a tagged photon in high-energy heavy ion collisions,''
  Phys.\ Rev.\ Lett.\  {\bf 77} (1996) 231
%  doi:10.1103/PhysRevLett.77.231
%  [hep-ph/9605213].
  %%CITATION = doi:10.1103/PhysRevLett.77.231;%%
  %262 citations counted in INSPIRE as of 05 Sep 2017

%\cite{Sjostrand:2007gs}
\bibitem{Sjostrand:2007gs}
  T.~Sj\"ostrand, S.~Mrenna and P.~Z.~Skands,
  %``A Brief Introduction to PYTHIA 8.1,''
  Comput.\ Phys.\ Commun.\  {\bf 178} (2008) 852.
%  doi:10.1016/j.cpc.2008.01.036
%  [arXiv:0710.3820 [hep-ph]].
  %%CITATION = doi:10.1016/j.cpc.2008.01.036;%%
  %3291 citations counted in INSPIRE as of 29 Aug 2017

%\cite{ALICE:2014rma}
\bibitem{ALICE:2014rma}
  B.~B.~Abelev {\it et al.} [ALICE Collaboration],
  %``Inclusive photon production at forward rapidities in proton-proton collisions at $\sqrt{s}$ = 0.9, 2.76 and 7 TeV,''
  Eur.\ Phys.\ J.\ C {\bf 75} (2015) 146.
%  doi:10.1140/epjc/s10052-015-3356-2
%  [arXiv:1411.4981 [nucl-ex]].
  %%CITATION = doi:10.1140/epjc/s10052-015-3356-2;%%
  %2 citations counted in INSPIRE as of 31 Aug 2017

%\cite{Adam:2015lda}
\bibitem{Adam:2015lda}
  J.~Adam {\it et al.} [ALICE Collaboration],
  %``Direct photon production in Pb-Pb collisions at $\sqrt{s_\rm{NN}} =$ 2.76 TeV,''
  Phys.\ Lett.\ B {\bf 754} (2016) 235.
%  doi:10.1016/j.physletb.2016.01.020
%  [arXiv:1509.07324 [nucl-ex]].
  %%CITATION = doi:10.1016/j.physletb.2016.01.020;%%
  %65 citations counted in INSPIRE as of 31 Aug 2017

%\cite{Aaboud:2017cbm}
\bibitem{Aaboud:2017cbm}
  M.~Aaboud {\it et al.} [ATLAS Collaboration],
  %``Measurement of the cross section for inclusive isolated-photon production in $pp$ collisions at $\sqrt s=13$ TeV using the ATLAS detector,''
  Phys.\ Lett.\ B {\bf 770} (2017) 473.
%  doi:10.1016/j.physletb.2017.04.072
%  [arXiv:1701.06882 [hep-ex]].
  %%CITATION = doi:10.1016/j.physletb.2017.04.072;%%
  %6 citations counted in INSPIRE as of 29 Aug 2017

%\cite{Aad:2015lcb}
\bibitem{Aad:2015lcb}
  G.~Aad {\it et al.} [ATLAS Collaboration],
  %``Centrality, rapidity and transverse momentum dependence of isolated prompt photon production in lead-lead collisions at $\sqrt{s_{\mathrm{NN}}} = 2.76$ TeV measured with the ATLAS detector,''
  Phys.\ Rev.\ C {\bf 93} (2016) 034914.
%  doi:10.1103/PhysRevC.93.034914
%  [arXiv:1506.08552 [hep-ex]].
  %%CITATION = doi:10.1103/PhysRevC.93.034914;%%
  %28 citations counted in INSPIRE as of 31 Aug 2017

%\cite{CMS:2013oua}
\bibitem{CMS:2013oua}
  CMS Collaboration [CMS Collaboration],
  %``Study of Isolated photon jet correlation in PbPb and pp collisions at 2.76TeV  and pPb collisions at 5.02TeV,''
  CMS-PAS-HIN-13-006.
  %%CITATION = CMS-PAS-HIN-13-006;%%
  %12 citations counted in INSPIRE as of 29 Aug 2017

%\cite{Aaij:2016ofv}
\bibitem{Aaij:2016ofv}
  R.~Aaij {\it et al.} [LHCb Collaboration],
  %``First experimental study of photon polarization in radiative $B^{0}_{s}$ decays,''
  Phys.\ Rev.\ Lett.\  {\bf 118} (2017)  021801
   Addendum: [Phys.\ Rev.\ Lett.\  {\bf 118} (2017)  109901].
%  doi:10.1103/PhysRevLett.118.021801, 10.1103/PhysRevLett.118.109901
%  [arXiv:1609.02032 [hep-ex]].
  %%CITATION = doi:10.1103/PhysRevLett.118.021801, 10.1103/PhysRevLett.118.109901;%%
  %10 citations counted in INSPIRE as of 05 Sep 2017

%\cite{Catani:2002ny}
\bibitem{Catani:2002ny}
  S.~Catani, M.~Fontannaz, J.~P.~Guillet and E.~Pilon,
  %``Cross-section of isolated prompt photons in hadron hadron collisions,''
  JHEP {\bf 0205} (2002) 028;
%  doi:10.1088/1126-6708/2002/05/028
%  [hep-ph/0204023].
  %%CITATION = doi:10.1088/1126-6708/2002/05/028;%%
  %209 citations counted in INSPIRE as of 14 Sep 2017
%
%\cite{Aurenche:2006vj}
%\bibitem{Aurenche:2006vj}
  P.~Aurenche, M.~Fontannaz, J.~P.~Guillet, E.~Pilon and M.~Werlen,
  %``A New critical study of photon production in hadronic collisions,''
  Phys.\ Rev.\ D {\bf 73} (2006) 094007.
%  doi:10.1103/PhysRevD.73.094007
%  [hep-ph/0602133].
  %%CITATION = doi:10.1103/PhysRevD.73.094007;%%
  %198 citations counted in INSPIRE as of 29 Aug 2017

%\cite{Aaltonen:2009ty}
\bibitem{Aaltonen:2009ty}
  T.~Aaltonen {\it et al.} [CDF Collaboration],
  %``Measurement of the Inclusive Isolated Prompt Photon Cross Section in $p\bar{p}$ Collisions at $\sqrt{s}=1.96$~TeV using the CDF Detector,''
  Phys.\ Rev.\ D {\bf 80} (2009) 111106.
%  doi:10.1103/PhysRevD.80.111106
%  [arXiv:0910.3623 [hep-ex]].
  %%CITATION = doi:10.1103/PhysRevD.80.111106;%%
  %53 citations counted in INSPIRE as of 16 Nov 2017

%\cite{Abazov:2013pua}
\bibitem{Abazov:2013pua}
  V.~M.~Abazov {\it et al.} [D0 Collaboration],
  %``Measurement of the differential cross sections for isolated direct photon pair production in $p \bar p$ collisions at $\sqrt{s} = 1.96$ TeV,''
  Phys.\ Lett.\ B {\bf 725} (2013) 6.
%  doi:10.1016/j.physletb.2013.06.036
%  [arXiv:1301.4536 [hep-ex]].
  %%CITATION = doi:10.1016/j.physletb.2013.06.036;%%
  %22 citations counted in INSPIRE as of 16 Nov 2017

%\cite{D0:2013lra}
\bibitem{D0:2013lra}
  V.~M.~Abazov {\it et al.} [D0 Collaboration],
  %``Measurement of the differential cross section of photon plus jet production in $p\bar{p}$ collisions at $\sqrt{s}=1.96$ TeV,''
  Phys.\ Rev.\ D {\bf 88} (2013) 072008.
%  doi:10.1103/PhysRevD.88.072008
%  [arXiv:1308.2708 [hep-ex]].
  %%CITATION = doi:10.1103/PhysRevD.88.072008;%%
  %14 citations counted in INSPIRE as of 16 Nov 2017

%\cite{Aaltonen:2013ama}
\bibitem{Aaltonen:2013ama}
  T.~Aaltonen {\it et al.} [CDF Collaboration],
  %``Measurement of the cross section for direct-photon production in association with a heavy quark in $p\bar{p}$ collisions at $\sqrt{s}$ = 1.96 TeV,''
  Phys.\ Rev.\ Lett.\  {\bf 111} (2013)  042003;
%  doi:10.1103/PhysRevLett.111.042003
%  [arXiv:1303.6136 [hep-ex]].
  %%CITATION = doi:10.1103/PhysRevLett.111.042003;%%
  %33 citations counted in INSPIRE as of 16 Nov 2017
%
%\cite{D0:2012gw}
%\bibitem{D0:2012gw}
  V.~M.~Abazov {\it et al.} [D0 Collaboration],
  %``Measurement of the differential photon + $c$-jet cross section and the ratio of differential photon+ $c$ and photon+ $b$ cross sections in proton-antiproton collisions at $\sqrt{s}=1.96$ TeV,''
  Phys.\ Lett.\ B {\bf 719} (2013) 354.
%  doi:10.1016/j.physletb.2013.01.033
%  [arXiv:1210.5033 [hep-ex]].
  %%CITATION = doi:10.1016/j.physletb.2013.01.033;%%
  %40 citations counted in INSPIRE as of 16 Nov 2017

%\cite{Bourhis:1997yu}
\bibitem{Bourhis:1997yu}
  L.~Bourhis, M.~Fontannaz and J.~P.~Guillet,
  %``Quarks and gluon fragmentation functions into photons,''
  Eur.\ Phys.\ J.\ C {\bf 2} (1998) 529.
%  doi:10.1007/s100520050158
%  [hep-ph/9704447].
  %%CITATION = doi:10.1007/s100520050158;%%
  %233 citations counted in INSPIRE as of 29 Aug 2017

%\cite{Klasen:2014xfa}
\bibitem{Klasen:2014xfa}
  M.~Klasen and F.~K\"onig,
  %``New information on photon fragmentation functions,''
  Eur.\ Phys.\ J.\ C {\bf 74} (2014) 3009;
%  doi:10.1140/epjc/s10052-014-3009-x
%  [arXiv:1403.2290 [hep-ph]].
  %%CITATION = doi:10.1140/epjc/s10052-014-3009-x;%%
  %8 citations counted in INSPIRE as of 05 Sep 2017
%
%\cite{Kaufmann:2016nux}
%\bibitem{Kaufmann:2016nux}
  T.~Kaufmann, A.~Mukherjee and W.~Vogelsang,
  %``Access to Photon Fragmentation Functions in Hadronic Jet Production,''
  Phys.\ Rev.\ D {\bf 93} (2016) 114021.
%  doi:10.1103/PhysRevD.93.114021
%  [arXiv:1604.07175 [hep-ph]].
  %%CITATION = doi:10.1103/PhysRevD.93.114021;%%
  %6 citations counted in INSPIRE as of 05 Sep 2017

%\cite{Campbell:2016yrh}
\bibitem{Campbell:2016yrh}
  J.~M.~Campbell, R.~K.~Ellis, Y.~Li and C.~Williams,
  %``Predictions for diphoton production at the LHC through NNLO in QCD,''
  JHEP {\bf 1607} (2016) 148;
%  doi:10.1007/JHEP07(2016)148
%  [arXiv:1603.02663 [hep-ph]].
  %%CITATION = doi:10.1007/JHEP07(2016)148;%%
  %40 citations counted in INSPIRE as of 16 Nov 2017
%
%\cite{Boughezal:2016wmq}
%\bibitem{Boughezal:2016wmq}
  R.~Boughezal, J.~M.~Campbell, R.~K.~Ellis, C.~Focke, W.~Giele, X.~Liu, F.~Petriello and C.~Williams,
  %``Color singlet production at NNLO in MCFM,''
  Eur.\ Phys.\ J.\ C {\bf 77} (2017)  7.
%  doi:10.1140/epjc/s10052-016-4558-y
%  [arXiv:1605.08011 [hep-ph]].
  %%CITATION = doi:10.1140/epjc/s10052-016-4558-y;%%
  %48 citations counted in INSPIRE as of 16 Nov 2017

%\cite{Campbell:2017aul}
\bibitem{Campbell:2017aul}
  J.~M.~Campbell, T.~Neumann and C.~Williams,
  %``$Z\gamma$ production at NNLO including anomalous couplings,''
  %Submitted to: JHEP
  [arXiv:1708.02925 [hep-ph]].
  %%CITATION = ARXIV:1708.02925;%%
  %3 citations counted in INSPIRE as of 16 Nov 2017

%\cite{Campbell:2012ft}
\bibitem{Campbell:2012ft}
  J.~M.~Campbell, H.~B.~Hartanto and C.~Williams,
  %``Next-to-leading order predictions for $Z \gamma$+jet and Z $\gamma \gamma$ final states at the LHC,''
  JHEP {\bf 1211} (2012) 162.
%  doi:10.1007/JHEP11(2012)162
%  [arXiv:1208.0566 [hep-ph]].
  %%CITATION = doi:10.1007/JHEP11(2012)162;%%
  %23 citations counted in INSPIRE as of 16 Nov 2017

%\cite{Frixione:1995ms}
\bibitem{Frixione:1995ms}
  S.~Frixione, Z.~Kunszt and A.~Signer,
  %``Three jet cross-sections to next-to-leading order,''
  Nucl.\ Phys.\ B {\bf 467} (1996) 399.
%  doi:10.1016/0550-3213(96)00110-1
%  [hep-ph/9512328].
  %%CITATION = doi:10.1016/0550-3213(96)00110-1;%%
  %623 citations counted in INSPIRE as of 01 Sep 2017

%\cite{Frixione:2007vw}
\bibitem{Frixione:2007vw}
  S.~Frixione, P.~Nason and C.~Oleari,
  %``Matching NLO QCD computations with Parton Shower simulations: the POWHEG method,''
  JHEP {\bf 0711} (2007) 070.
%  doi:10.1088/1126-6708/2007/11/070
%  [arXiv:0709.2092 [hep-ph]].
  %%CITATION = doi:10.1088/1126-6708/2007/11/070;%%
  %2026 citations counted in INSPIRE as of 01 Sep 2017

%\cite{Jezo:2016ypn}
\bibitem{Jezo:2016ypn}
  T.~Jezo, M.~Klasen and F.~K\"onig,
  %``Prompt photon production and photon-hadron jet correlations with POWHEG,''
  JHEP {\bf 1611} (2016) 033.
%  doi:10.1007/JHEP11(2016)033
%  [arXiv:1610.02275 [hep-ph]].
  %%CITATION = doi:10.1007/JHEP11(2016)033;%%
  %5 citations counted in INSPIRE as of 29 Aug 2017

%\cite{Hoeche:2009xc}
\bibitem{Hoeche:2009xc}
  S.~H\"oche, S.~Schumann and F.~Siegert,
  %``Hard photon production and matrix-element parton-shower merging,''
  Phys.\ Rev.\ D {\bf 81} (2010) 034026.
%  doi:10.1103/PhysRevD.81.034026
%  [arXiv:0912.3501 [hep-ph]].
  %%CITATION = doi:10.1103/PhysRevD.81.034026;%%
  %101 citations counted in INSPIRE as of 05 Sep 2017

%\cite{Gehrmann:2013bga}
\bibitem{Gehrmann:2013bga}
  T.~Gehrmann, N.~Greiner and G.~Heinrich,
  %``Precise QCD predictions for the production of a photon pair in association with two jets,''
  Phys.\ Rev.\ Lett.\  {\bf 111} (2013) 222002;
%  doi:10.1103/PhysRevLett.111.222002
%  [arXiv:1308.3660 [hep-ph]].
  %%CITATION = doi:10.1103/PhysRevLett.111.222002;%%
  %50 citations counted in INSPIRE as of 23 Nov 2017
%
%\cite{Badger:2013ava}
%\bibitem{Badger:2013ava}
  S.~Badger, A.~Guffanti and V.~Yundin,
  %``Next-to-leading order QCD corrections to di-photon production in association with up to three jets at the Large Hadron Collider,''
  JHEP {\bf 1403} (2014) 122;
%  doi:10.1007/JHEP03(2014)122
%  [arXiv:1312.5927 [hep-ph]].
  %%CITATION = doi:10.1007/JHEP03(2014)122;%%
  %32 citations counted in INSPIRE as of 23 Nov 2017
%
%\cite{Bern:2014vza}
%\bibitem{Bern:2014vza}
  Z.~Bern, L.~J.~Dixon, F.~Febres Cordero, S.~Hoeche, H.~Ita, D.~A.~Kosower, N.~A.~Lo Presti and D.~Maitre,
  %``Next-to-leading order $\gamma \gamma+2$-jet production at the LHC,''
  Phys.\ Rev.\ D {\bf 90} (2014) 054004.
%  doi:10.1103/PhysRevD.90.054004
%  [arXiv:1402.4127 [hep-ph]].
  %%CITATION = doi:10.1103/PhysRevD.90.054004;%%
  %24 citations counted in INSPIRE as of 23 Nov 2017

%\cite{Siegert:2016bre}
\bibitem{Siegert:2016bre}
  F.~Siegert,
  %``A practical guide to event generation for prompt photon production with Sherpa,''
  J.\ Phys.\ G {\bf 44} (2017) 044007.
%  doi:10.1088/1361-6471/aa5f29
%  [arXiv:1611.07226 [hep-ph]].
  %%CITATION = doi:10.1088/1361-6471/aa5f29;%%
  %6 citations counted in INSPIRE as of 17 Jan 2018

%\cite{Fritzsche:2013fta}
\bibitem{Fritzsche:2013fta}
  T.~Fritzsche, T.~Hahn, S.~Heinemeyer, F.~von der Pahlen, H.~Rzehak and C.~Schappacher,
  %``The Implementation of the Renormalized Complex MSSM in FeynArts and FormCalc,''
  Comput.\ Phys.\ Commun.\  {\bf 185} (2014) 1529.
%  doi:10.1016/j.cpc.2014.02.005
%  [arXiv:1309.1692 [hep-ph]].
  %%CITATION = doi:10.1016/j.cpc.2014.02.005;%%
  %33 citations counted in INSPIRE as of 01 Sep 2017

%\cite{Alwall:2011uj}
\bibitem{Alwall:2011uj}
  J.~Alwall, M.~Herquet, F.~Maltoni, O.~Mattelaer and T.~Stelzer,
  %``MadGraph 5 : Going Beyond,''
  JHEP {\bf 1106} (2011) 128.
%  doi:10.1007/JHEP06(2011)128
%  [arXiv:1106.0522 [hep-ph]].
  %%CITATION = doi:10.1007/JHEP06(2011)128;%%
  %2460 citations counted in INSPIRE as of 01 Sep 2017

%\cite{Vermaseren:2000nd}
\bibitem{Vermaseren:2000nd}
  J.~A.~M.~Vermaseren,
  %``New features of FORM,''
  math-ph/0010025.
  %%CITATION = MATH-PH/0010025;%%
  %1294 citations counted in INSPIRE as of 01 Sep 2017

%\cite{Hahn:1998yk}
\bibitem{Hahn:1998yk}
  T.~Hahn and M.~Perez-Victoria,
  %``Automatized one loop calculations in four-dimensions and D-dimensions,''
  Comput.\ Phys.\ Commun.\  {\bf 118} (1999) 153.
%  doi:10.1016/S0010-4655(98)00173-8
%  [hep-ph/9807565].
  %%CITATION = doi:10.1016/S0010-4655(98)00173-8;%%
  %1314 citations counted in INSPIRE as of 01 Sep 2017

%\cite{Alwall:2014hca}
\bibitem{Alwall:2014hca}
  J.~Alwall {\it et al.},
  %``The automated computation of tree-level and next-to-leading order differential cross sections, and their matching to parton shower simulations,''
  JHEP {\bf 1407} (2014) 079.
%  doi:10.1007/JHEP07(2014)079
%  [arXiv:1405.0301 [hep-ph]].
  %%CITATION = doi:10.1007/JHEP07(2014)079;%%
  %2120 citations counted in INSPIRE as of 01 Sep 2017

%\cite{Catani:1996vz}
\bibitem{Catani:1996vz}
  S.~Catani and M.~H.~Seymour,
  %``A General algorithm for calculating jet cross-sections in NLO QCD,''
  Nucl.\ Phys.\ B {\bf 485} (1997) 291
   Erratum: [Nucl.\ Phys.\ B {\bf 510} (1998) 503].
%  doi:10.1016/S0550-3213(96)00589-5, 10.1016/S0550-3213(98)81022-5
%  [hep-ph/9605323].
  %%CITATION = doi:10.1016/S0550-3213(96)00589-5, 10.1016/S0550-3213(98)81022-5;%%
  %1460 citations counted in INSPIRE as of 01 Sep 2017

%\cite{Hasegawa:2009tx}
\bibitem{Hasegawa:2009tx}
  K.~Hasegawa, S.~Moch and P.~Uwer,
  %``AutoDipole: Automated generation of dipole subtraction terms,''
  Comput.\ Phys.\ Commun.\  {\bf 181} (2010) 1802.
%  doi:10.1016/j.cpc.2010.06.044
%  [arXiv:0911.4371 [hep-ph]].
  %%CITATION = doi:10.1016/j.cpc.2010.06.044;%%
  %50 citations counted in INSPIRE as of 01 Sep 2017

%\cite{DErrico:2011cgc}
\bibitem{DErrico:2011cgc}
  L.~D'Errico and P.~Richardson,
  %``Next-to-Leading-Order Monte Carlo Simulation of Diphoton Production in Hadronic Collisions,''
  JHEP {\bf 1202} (2012) 130.
%  doi:10.1007/JHEP02(2012)130
%  [arXiv:1106.3939 [hep-ph]].
  %%CITATION = doi:10.1007/JHEP02(2012)130;%%
  %23 citations counted in INSPIRE as of 16 Nov 2017

%\cite{Barze:2014zba}
\bibitem{Barze:2014zba}
  L.~Barze, M.~Chiesa, G.~Montagna, P.~Nason, O.~Nicrosini, F.~Piccinini and V.~Prosperi,
  %``W$\gamma$ production in hadronic collisions using the POWHEG+MiNLO method,''
  JHEP {\bf 1412} (2014) 039.
%  doi:10.1007/JHEP12(2014)039
%  [arXiv:1408.5766 [hep-ph]].
  %%CITATION = doi:10.1007/JHEP12(2014)039;%%
  %12 citations counted in INSPIRE as of 16 Nov 2017

%\cite{Ball:2017nwa}
\bibitem{Ball:2017nwa}
  R.~D.~Ball {\it et al.} [NNPDF Collaboration],
  %``Parton distributions from high-precision collider data,''
  arXiv:1706.00428 [hep-ph].
  %%CITATION = ARXIV:1706.00428;%%
  %14 citations counted in INSPIRE as of 29 Aug 2017

%\cite{Ball:2013hta}
\bibitem{Ball:2013hta}
  R.~D.~Ball {\it et al.} [NNPDF Collaboration],
  %``Parton distributions with QED corrections,''
  Nucl.\ Phys.\ B {\bf 877} (2013) 290.
%  doi:10.1016/j.nuclphysb.2013.10.010
%  [arXiv:1308.0598 [hep-ph]].
  %%CITATION = doi:10.1016/j.nuclphysb.2013.10.010;%%
  %257 citations counted in INSPIRE as of 29 Aug 2017

%\cite{Skands:2014pea}
\bibitem{Skands:2014pea}
  P.~Skands, S.~Carrazza and J.~Rojo,
  %``Tuning PYTHIA 8.1: the Monash 2013 Tune,''
  Eur.\ Phys.\ J.\ C {\bf 74} (2014) 3024.
%  doi:10.1140/epjc/s10052-014-3024-y
%  [arXiv:1404.5630 [hep-ph]].
  %%CITATION = doi:10.1140/epjc/s10052-014-3024-y;%%
  %227 citations counted in INSPIRE as of 29 Aug 2017

\bibitem{cacciari}
 M.\ Cacciari, D.\ d'Enterria, private communication.

%\cite{Cacciari:2008gp}
\bibitem{Cacciari:2008gp}
  M.~Cacciari, G.~P.~Salam and G.~Soyez,
  %``The Anti-k(t) jet clustering algorithm,''
  JHEP {\bf 0804} (2008) 063.
%  doi:10.1088/1126-6708/2008/04/063
%  [arXiv:0802.1189 [hep-ph]].
  %%CITATION = doi:10.1088/1126-6708/2008/04/063;%%
  %4609 citations counted in INSPIRE as of 29 Aug 2017

%\cite{Wang:1991hta}
\bibitem{Wang:1991hta}
  X.~N.~Wang and M.~Gyulassy,
  %``HIJING: A Monte Carlo model for multiple jet production in p p, p A and A A collisions,''
  Phys.\ Rev.\ D {\bf 44} (1991) 3501.
  doi:10.1103/PhysRevD.44.3501
  %%CITATION = doi:10.1103/PhysRevD.44.3501;%%
  %1380 citations counted in INSPIRE as of 30 Aug 2017

%\cite{CMS:2016ynj}
\bibitem{CMS:2016ynj}
  CMS Collaboration [CMS Collaboration],
  %``Study of Isolated-Photon + Jet Correlations in PbPb and pp Collisions at $\sqrt{s_{NN}} = 5.02\:\mathrm{TeV}$,''
  CMS-PAS-HIN-16-002.
  %%CITATION = CMS-PAS-HIN-16-002;%%
  %4 citations counted in INSPIRE as of 31 Aug 2017

%\cite{ATLAS:2017gxj}
\bibitem{ATLAS:2017gxj}
  The ATLAS collaboration [ATLAS Collaboration],
  %``Measurement of the cross section for isolated-photon plus jet production in $pp$ collisions at $\sqrt s=13$ TeV using the ATLAS detector,''
  ATLAS-CONF-2017-059.
  %%CITATION = ATLAS-CONF-2017-059;%%

%\cite{Kovarik:2015cma}
\bibitem{Kovarik:2015cma}
  K.~Kovarik {\it et al.},
  %``nCTEQ15 - Global analysis of nuclear parton distributions with uncertainties in the CTEQ framework,''
  Phys.\ Rev.\ D {\bf 93} (2016) 085037.
%  doi:10.1103/PhysRevD.93.085037
%  [arXiv:1509.00792 [hep-ph]].
  %%CITATION = doi:10.1103/PhysRevD.93.085037;%%
  %76 citations counted in INSPIRE as of 08 Sep 2017

%\cite{Stump:2003yu}
\bibitem{Stump:2003yu}
  D.~Stump, J.~Huston, J.~Pumplin, W.~K.~Tung, H.~L.~Lai, S.~Kuhlmann and J.~F.~Owens,
  %``Inclusive jet production, parton distributions, and the search for new physics,''
  JHEP {\bf 0310} (2003) 046.
%  doi:10.1088/1126-6708/2003/10/046
%  [hep-ph/0303013].
  %%CITATION = doi:10.1088/1126-6708/2003/10/046;%%
  %882 citations counted in INSPIRE as of 08 Sep 2017

%\cite{Eskola:2016oht}
\bibitem{Eskola:2016oht}
  K.~J.~Eskola, P.~Paakkinen, H.~Paukkunen and C.~A.~Salgado,
  %``EPPS16: Nuclear parton distributions with LHC data,''
  Eur.\ Phys.\ J.\ C {\bf 77} (2017) 163.
%  doi:10.1140/epjc/s10052-017-4725-9
%  [arXiv:1612.05741 [hep-ph]].
  %%CITATION = doi:10.1140/epjc/s10052-017-4725-9;%%
  %24 citations counted in INSPIRE as of 08 Sep 2017

%\cite{CMS:2015hha}
\bibitem{CMS:2015hha}
  CMS Collaboration [CMS Collaboration],
  %``Production of pairs of isolated photons in association with jets in pp collisions at sqrt(s) = 7 TeV,''
  CMS-PAS-SMP-14-021.
  %%CITATION = CMS-PAS-SMP-14-021;%%
  %6 citations counted in INSPIRE as of 11 Sep 2017

%\cite{germain}
\bibitem{germain}
  M.~Germain [ALICE Collaboration],
  %``Direct photon measurements in pp and Pb-Pb collisions with the ALICE experiment,''
  Proceedings of the XXVI International Conference on Ultrarelativistic Heavy-Ion Collisions
  (Quark Matter 2017), Chicago, USA.
%  Nucl.\ Phys.\ A, to be published.

%\cite{Adam:2015hoa}
\bibitem{Adam:2015hoa}
  J.~Adam {\it et al.} [ALICE Collaboration],
  %``Measurement of charged jet production cross sections and nuclear modification in p-Pb collisions at $\sqrt{s_\rm{NN}} = 5.02$ TeV,''
  Phys.\ Lett.\ B {\bf 749} (2015) 68
%  doi:10.1016/j.physletb.2015.07.054
%  [arXiv:1503.00681 [nucl-ex]].
  %%CITATION = doi:10.1016/j.physletb.2015.07.054;%%
  %37 citations counted in INSPIRE as of 11 Sep 2017
%
%\cite{Adam:2016jfp}
%\bibitem{Adam:2016jfp}
  J.~Adam {\it et al.} [ALICE Collaboration],
  %``Centrality dependence of charged jet production in p–Pb collisions at $\sqrt{s_\mathrm{NN}}$ = 5.02 TeV,''
  Eur.\ Phys.\ J.\ C {\bf 76} (2016) 271.
%  doi:10.1140/epjc/s10052-016-4107-8
%  [arXiv:1603.03402 [nucl-ex]].
  %%CITATION = doi:10.1140/epjc/s10052-016-4107-8;%%
  %14 citations counted in INSPIRE as of 11 Sep 2017

%\cite{AbellanBeteta:2016ugk}
\bibitem{AbellanBeteta:2016ugk}
  R.~Aaij {\it et al.} [LHCb Collaboration],
  %``Measurement of forward $W$ and $Z$ boson production in association with jets in proton-proton collisions at $\sqrt{s}=8$ TeV,''
  JHEP {\bf 1605} (2016) 131.
%  doi:10.1007/JHEP05(2016)131
%  [arXiv:1605.00951 [hep-ex]].
  %%CITATION = doi:10.1007/JHEP05(2016)131;%%
  %8 citations counted in INSPIRE as of 11 Sep 2017

\end{thebibliography}

\end{document}